\documentclass[AMA,Times1COL]{WileyNJDv5} 

\articletype{Article Type}%

\received{Date Month Year}
\revised{Date Month Year}
\accepted{Date Month Year}
\journal{Journal}
\volume{00}
\copyyear{2023}
\startpage{1}
\usepackage{lineno}
\usepackage{adjustbox}
\usepackage{dsfont}
\usepackage{multirow}
\usepackage{caption}
\usepackage{placeins}
\usepackage{todonotes}

\begin{document}

\title{Scalable Gaussian Process for Learning Non-Ergodic Ground Motion Model from Physics-Based Simulations with Application to Power Infrastructure Assessment}

\author[1]{Jinyan Zhao}

\author[2]{Grigorios Lavrentiadis}

\author[1]{Domniki Asimaki}

\authormark{Zhao \textsc{et al.}}
\titlemark{Scalable Gaussian Process for Learning Non-Ergodic Ground Motion Model from Physics-Based Simulations and Application to Power Infrastructure Assessment}

\address[1]{\orgdiv{Mechanical and Civil Engineering}, \orgname{California Institute of Technology}, \orgaddress{\state{CA}, \country{USA}}}

\address[2]{\orgdiv{Dept. of Civil, Structural and Environmental Engineering}, \orgname{State University of New York at Buffalo}, \orgaddress{\state{NY}, \country{USA}}}


\corres{Corresponding author Grigorios Lavrentiadis. \email{glavrent@buffalo.edu} }



\abstract[Abstract]{This study presents the development and application of a scalable non-ergodic ground motion model (NGMM) for the Los Angeles area. The NGMM is trained and validated on physics-based simulated ground-motion data from a recent Statewide California Earthquake Center (SCEC) CyberShake study. The NGMM is formulated as a Gaussian Process (GP) regression model, where the prior median is defined as the ASK14 ergodic ground-motion model and the posterior median is obtained by learning the non-ergodic effects embedded in the training data. These non-ergodic effects include systematic site and path effects, which are represented in the GP using Matérn and specialized covariance kernels that explicitly characterize path vectors. Implementing the NGMM requires hyperparameter tuning and inference on large datasets (on the order of one million data points or more), posing significant computational challenges for conventional GP approaches. To address this scalability issue, this paper presents a suite of computational strategies, including sparse Cholesky inversion, parallel computing, GPU acceleration, and stochastic gradient descent minimization. Despite these advances, the full CyberShake dataset (on the order of hundreds of millions of data points) remains computationally prohibitive. Therefore, aleatory variability is modeled separately using a mixed-effects formulation to represent within-event and between-event variability.
The developed NGMM has two primary applications: interpolation of partially observed ground-motion fields and predictive modeling for ground motions in unobserved earthquake scenarios. Validation results on independent datasets demonstrate accurate performance in both applications. A case study of power transmission network performance assessment in an Mw 6.7 Puente Hill scenario further demonstrated that the developed NGMM closely reproduces physics-based simulation results at a substantially reduced computational cost, and that neglecting non-ergodic effects in ground motion modeling can bias damage estimates by nearly a factor of two.}


\jnlcitation{\cname{%
\author{Zhao J.},
\author{Lavrentiadis G.}, and
\author{Asimaki D.}}.
\ctitle{} \cjournal{\it Earthquake Engineering and Structural Dynamics} \cvol{2026;00(00):1--18}.}

\maketitle




\section{Introduction}\label{sec:introduction}

Conventional ground motion models (GMMs) use multivariate lognormal (MVN) distributions to represent aleatory variability, i.e., the discrepancy between recorded ground-motion intensity measures and GMM median predictions. The covariance structure of the MVN distribution is typically decomposed into within-event and between-event components \cite{abrahamson2014summary,baker2017intensity,markhvida2018modeling}, with covariance functions estimated from ground-motion datasets collected across similar tectonic environments. The rationale for combining data in similar tectonic environments is the ergodic assumption \cite{anderson1999probabilistic}, which pulls together available data from similar tectonic environments to inform ground motion scaling in the GMMs. Although the ergodic assumption facilitates robust estimation of general trends in ground motion predictions, it also introduces large aleatory variability by neglecting region-specific effects. 

The growing availability of modern ground-motion recordings has motivated the development of non-ergodic ground motion models (NGMMs), which explicitly capture region-specific effects and reduce predictive variability. A widely adopted approach for NGMMs is Gaussian Process (GP) regression, in which an ergodic GMM serves as the backbone (prior median), and the posterior distribution is obtained by conditioning on observed data through a Gaussian likelihood. The covariance of the Gaussian likelihood is defined via a kernel function, which is commonly constructed as a sum of source, site, and path kernels in log space to represent systematic contributions from earthquake sources, site conditions, and wave propagation paths, respectively. By conditioning on region-specific ground motion records, GP-based NGMMs infer systematic non-ergodic effects and thereby reduce predictive variability. A comprehensive review of non-ergodic effects and GP formulations for NGMMs can be found in Lavrentiadis et al. \cite{lavrentiadis2023overview}. 

Considering the requirement for ground motion data to infer repeatable effects,  physics-based ground motion simulations have emerged as a valuable supplementary source in the NGMM development \cite{Lavrentiadis2024Groningen}. Physics-based simulations generate earthquake ground motions by explicitly modeling earthquakes as ruptures on the causative faults and the consequent seismic wave propagation using momentum conservation principles. Baker et al. \cite{baker2021physics} provide an elegant introduction to physics-based ground motion simulations. Although subject to various uncertainties, such as modeling the fault-rupturing process and crustal velocity structures, physics-based simulations offer densely sampled ground motions, particularly in near-source regions where observational data are sparse, thereby offering valuable insights and potential engineering applications \cite{baker2014engineering}. However, physics-based ground-motion simulations are computationally intensive, which limits their use in seismic hazard and risk analysis. In response to this, this paper develops an NGMM using data generated in the Statewide California Earthquake Center (SCEC) CyberShake Study 22.12, a carefully designed and validated physics-based ground-motion simulation dataset. The developed NGMM has two primary applications. First, for earthquake scenarios simulated in CyberShake, it enables interpolation of ground motions at unrecorded locations. Second, for scenarios outside the CyberShake catalog, it serves as a surrogate model for physics-based simulations, enabling efficient prediction of ground-motion intensities at arbitrary locations within the study region. In both cases, the NGMM provides rigorous uncertainty quantification. Although being more accurate than ergodic GMMs and less computationally intensive than physics-based simulations, GP-based NGMMs can still be computationally demanding when applied to large datasets. For example, the CyberShake Study 22.12 dataset used in this study contains over 200 million data points. Consequently, a scalable NGMM implementation is developed in this paper to overcome this challenge. 

The remainder of this paper is organized as follows. We first describe the training dataset and the formulation of the NGMM, followed by the training and inference strategies that enable efficient hyperparameter tuning and prediction. We then demonstrate the advantages of the proposed NGMM over conventional ergodic GMMs by comparing predicted ground motions and probabilistic seismic hazard estimations against those obtained from the backbone GMM and physics-based simulations. Finally, a case study of seismic performance analysis for a synthetic power transmission network in Southern California under an Mw 6.7 Puente Hills scenario is presented. The case study illustrates the two application scenarios of the developed NGMM and highlights the importance of accounting for non-ergodic effects to achieve realistic seismic performance assessments.

\section{Background}
\subsection{Training data: SCEC CyberShake Study 22.12}
The dataset used in this study is Study 22.12 \cite{cybershake_study_22_12}, produced by CyberShake \cite{graves2011cybershake}, a high-performance computational platform developed by the SCEC. CyberShake is an integrated suite of scientific software that performs three-dimensional physics-based kinematic rupture and wave-propagation simulations. The CyberShake team continuously refines the underlying models and releases datasets from simulation studies of physics-based probabilistic seismic hazard in California. In this work, we utilize data from one of the most recent studies, Study 22.12.

Study 22.12 simulates seismograms at 335 sites across Southern California (see Fig.~\ref{fig:training_testing_sites_map}). The earthquake sources are derived from a catalog of fault-rupture scenarios based on the Second Uniform California Earthquake Rupture Forecast  (UCERF2)\cite{field2009uniform}, enhanced with variations in hypocenter locations and slip distributions. The catalog includes 9,283 rupture scenarios, and for each scenario, an average of 84 variations in hypocenter and slip distribution are generated using a kinematic stochastic rupture generator \cite{pitarka2022refinements}, resulting in more than 770,000 rupture variations. For each rupture variation, seismograms are generated using a hybrid simulation approach. The low-frequency (0–1 Hz) component is computed using a parallelized algorithm that is based on a set of strain Green tensors generated with a finite-difference method  \cite{graves1996simulating,graves2001resolution}, while the high-frequency (1–50 Hz) component is simulated with a stochastic model and merged with the low-frequency results following the approach of Graves and Pitarka \cite{graves2010broadband,graves2015refinements}. Accurate wave propagation is simulated through a detailed crustal velocity model that combines CVM-S4.26.M01 \cite{cvm_s4_26_m01_scecpedia} for deeper subsurface layers with the Ely–Jordan geotechnical layer (GTL) \cite{ely2010vs30} near the surface. From the simulated seismograms, pseudo-spectral accelerations (PSA) at 44 periods are extracted. For each site, only rupture variations within 200 km are retained, resulting in an average of approximately 626,000 rupture variations per site and over 200 million seismograms in total. The performed simulations required 728,832.4 node-hours and 2,330.9 wall-clock hours on high-performance computing systems. Additional details on CyberShake and Study 22.12 are available on SCEC web pages \cite{scec_cybershake_software, cybershake_study_22_12}.

In this paper, a few earthquake scenarios are excluded due to inconsistencies between the rupture geometries and reported distance metrics \citep{MaechlingCallaghanPersonalCommunication}. As a result, the final dataset consists of 206.7 million ground-motion records from 8,358 rupture scenarios and approximately 700,000 rupture variations. Figure~\ref{fig:training_testing_earthquakes_map} illustrates the geometry, magnitudes, and annual occurrence rates of the earthquake scenarios considered. Because the high-frequency component of the simulated seismograms is generated using a semi-stochastic approach that does not capture region-specific non-ergodic effects, the NGMM in this study is developed for PSA at a period of 2.0~s.

\subsection{NGMM formulation} \label{sec:NGMM_formulation}
The proposed NGMM formulation extends the work proposed in Lavrentiadis et al. \cite{Lavrentiadis2024Groningen}. In this framework, the logarithm of pseudo-spectral acceleration (PSA) at a period of 2.0 s, induced by an earthquake $e$ at a site $s$ is expressed as Eq. \ref{eq:NGMM_formulation_1}, where $f_{erg}(M, R_{rup}, V_{s,30}, ... )$ denotes an ergodic backbone ground motion model, $\delta P2P_{es}$ and $\delta S2S_{s}$ represent the non-ergodic path and site effects, and $\delta B^0_e$ and $\delta W^0_{es}$  denote the remaining between-event and within-event aleatory variability. In this study, the ASK14 model \cite{abrahamson2014summary} is used as the backbone model, which captures the scaling of ground motion against earthquake magnitude ($Mw$), the distance between earthquake rupture and site ($R_{rup}$), the time-averaged top 30 m shear velocity ($V_{s,30}$), and several other mechanisms. In the NGMM literature, the terms $\delta P2P_{es}$ and $\delta S2S_{s}$ are interpreted as repetitive effects due to systematic wave propagation and site amplification effects that can be determined from available data. The magnitude of these terms characterizes the epistemic uncertainty of the developed model.  In contrast, $\delta B^0_e$ and $\delta W^0_{es}$ represent the aleatory variability, which can not be reduced within the NGMM modeling framework, even with additional observations. Typical NGMMs in the literature also include an additional epistemic term to account for the repeatable source effects ($\delta L2L$). Such a source effect is not included in this study, because all sources in CyberShake are generated with the same rupture generator, and it does not introduce any source-specific repeatable effects. Gaussian Processes (GPs) are used in NGMM to model and capture the non-ergodic effects. For computational convenience, GPs are typically developed for the residual $y_{es}$, which is defined as the difference between $\mathrm{ln(PSA)}$ and the ergodic ground motion model, as shown in Eq. \ref{eq:NGMM_formulation_2}.
\begin{subequations} \label{eq:NGMM_formulation}
\begin{align}
    & \mathrm{ln(PSA)} = f_{erg}(Mw, R_{rup}, V_{s,30}, ... ) + \delta P2P_{es} + \delta S2S_{s} + \delta B^0_e + \delta W^0_{es}  \label{eq:NGMM_formulation_1} \\
    & y_{es} = \mathrm{ln(PSA)}  - f_{erg}(Mw, R_{rup}, V_{s30}, ... ) = \delta P2P_{es} + \delta S2S_{s} + \delta B^0_e + \delta W^0_{es} \label{eq:NGMM_formulation_2}
\end{align}
\end{subequations}

While taking $f_{erg}(Mw, R_{rup}, V_{s30}, ... )$ as the GP prior, the joint distribution of observed residual ($\mathbf{y}_{\mathrm{obs}}$) and the predicted mean residual ($\mathbf{\overline{y}}_{\mathrm{pred}}$) follows a multivariate normal distribution characterized with zero mean and a covariance matrix, as shown in Eq. \ref{eq:GP_formulation_1}. Here, boldface symbols denote vectors or matrices. $\mathbf{y}_{\mathrm{obs}}$ is a vector representing a collection of residuals across earthquake-site pairs in the conditional (training) dataset, while $\mathbf{\overline{y}}_{\mathrm{pred}}$ is a vector of the mean residual (adjustment) for the prediction scenarios. $\mathbf{K}_{f}$, $\mathbf{k}_{f}$, and $\mathbf{K}^{*}_{f}$ are blocks in the covariance matrix that are evaluated using a kernel function. The kernel function defined in \cite{Lavrentiadis2024Groningen} has been shown to
effectively capture the correlation structure of non-ergodic site and path effects and is adopted in this paper. The terms $\phi_0^{2}\mathbf{I}$ and $\tau_0^{2}\mathbf{1}$ represent the covariance matrices for the remaining within-event and between-event variability, where $\mathbf{I}$ is an identity matrix, and  $\mathbf{1}$ is a block matrix with entries
equal to 1 for pairs of observations from the same earthquake and 0 otherwise. In other words, the value $\delta B^0_e$ is fully correlated within the same event, and it has a single index $e$. 

\begin{subequations} \label{eq:GP_formulation}
\begin{align}
    & \begin{bmatrix}
        \mathbf{y}_{\mathrm{obs}} \\
        \mathbf{\overline{y}}_{\mathrm{pred}}
        \end{bmatrix}
        \sim
        \mathcal{N}\!\left(
        \mathbf{0},
        \begin{bmatrix}
        \mathbf{K}_{f} + \phi_0^{2}\mathbf{I} + \tau_0^{2}\mathbf{1} & \mathbf{k}_{f} \\
        \mathbf{k}_{f}^{\mathsf{T}} & \mathbf{K}^{*}_{f}
        \end{bmatrix}
        \right) \label{eq:GP_formulation_1}\\
    & \mathbb{E}\left[\mathbf{\overline{y}}_{\mathrm{pred}}|\mathbf{y}_{\mathrm{obs}}\right] = \mathbf{k}_{f} \left[ \mathbf{K}_{f} + \phi_0^{2}\mathbf{I} + \tau_0^{2}\mathbf{1} \right]^{-1} \mathbf{y}_{\mathrm{obs}} \label{eq:GP_formulation_2}\\
    & \mathrm{COV}\left[\mathbf{\overline{y}}_{\mathrm{pred}}|\mathbf{y}_{\mathrm{obs}}\right] = \mathbf{K}^{*}_{f} - \mathbf{k}_{f}^{\mathsf{T}} \left[ \mathbf{K}_{f} + \phi_0^{2}\mathbf{I} + \tau_0^{2}\mathbf{1} \right]^{-1}\mathbf{k}_{f} + \phi_0^{2}\mathbf{I} + \tau_0^{2}\mathbf{1} \label{eq:GP_formulation_3}
\end{align}
\end{subequations} 

While making inference with GP regression, the posterior distribution of $\mathbf{y}_{\mathrm{pred}}$ remains Gaussian, with mean and covariance given by Eq. \ref{eq:GP_formulation_2} and \ref{eq:GP_formulation_3}, respectively. Because the covariance matrix $\left[ \mathbf{K}_{f} + \phi_0^{2}\mathbf{I} + \tau_0^{2}\mathbf{1} \right]$ is positive definite, the posterior variance in Eq. \ref{eq:GP_formulation_3} is generally less than the prior variance $\mathbf{K}^*_{f}$. The reduction in variance reflects the ability of NGMMs to decrease
predictive uncertainty relative to ergodic GMMs by incorporating regional data. Before using the GP for inference, the hyperparameters in the kernel function need to be calibrated to reflect the characteristics of ground motion in the region of interest. This calibration is
typically performed using maximum likelihood estimation, in which the hyperparameters are determined by minimizing the negative log-likelihood defined with Eq. \ref{eq:loglikelihood}. Because of the complexity of the kernel function, closed-form solutions are generally not available, and gradient-based numerical optimization methods are employed. Further details on GP regression can be found in Williams and Rasmussen \cite{williams2006gaussian}.
\begin{equation} \label{eq:loglikelihood}
\ln p\!\left(
\mathbf{y}_{\mathrm{obs}} \mid 
\mathbf{x}_{\mathrm{obs}}
\right)
=
-\frac{1}{2}\,
\mathbf{y}_{\mathrm{obs}}^{\mathsf{T}}
\left(
\mathbf{K}_{f} + \phi_0^{2}\mathbf{I} + \tau_0^{2}\mathbf{I}
\right)^{-1}
\mathbf{y}_{\mathrm{obs}}
-\frac{1}{2}
\log\!\left|
\mathbf{K}_{f} + \phi_0^{2}\mathbf{I} + \tau_0^{2}\mathbf{I}
\right|
-\text{constant}
\end{equation}

\section{Training and inference strategy}
\subsection{Collocation data points and linear mixed model} \label{sec:collocation}

Different from NGMMs trained on recorded ground motions, the CyberShake simulations provide multiple rupture variations for each rupture scenario. These variations represent a range of possible fault slip distributions that share the same rupture geometry and magnitude. The rupture variations describe the aleatory variability in the fault rupture process related to inherent randomness, such as fault surface heterogeneity, and the recorded earthquakes can be viewed as an individual realizations of such a stochastic rupture process. Training the NGMM with physics-based simulations therefore enables direct observation and quantification of this source of aleatory variability. 

Because the kernel functions depend only on site coordinates and rupture locations, the kernel values evaluated for different rupture variations within the same scenario are identical. These data points with identical input variables but different response values are referred to as \emph{collocation data points}. While collocation points do not invalidate the GP formulation described in Section~\ref{sec:NGMM_formulation}, they can substantially increase the size of the covariance matrices, leading to prohibitive computational costs. The computational demand is particularly critical for this study, where the dataset contains hundreds of millions of data points. To address this challenge, aleatory variability terms ($\delta B^0_e $ and $ \delta W^0_{es}$) are decomposed into two components, as shown in Eq. \ref{eq:two_steps}. 

\begin{subequations} \label{eq:two_steps}
\begin{gather}
    \delta B_e  = \delta \dot{B}_l + \delta \ddot{B}_e\label{eq:two_steps_1} \\
    \delta W_{es} = \delta \dot{W}_{ls} + \delta \ddot{W}_{es} \label{eq:two_steps_2} \\
    \begin{aligned} \label{eq:two_steps_3}
    y_{es} &= \delta P2P_{ls} + \delta S2S_{s} + \delta \dot{B}_l  + \delta \dot{W}_{ls} + \delta \ddot{B}_e + \delta \ddot{W}_{es}  \\
     & = \bar{y}_{ls} + \delta \ddot{B}_e + \delta \ddot{W}_{es}
    \end{aligned}
\end{gather}
\end{subequations}
where $s$ is the site index representing different site locations, $l$ is the index representing different rupture scenarios in UCERF2 with distinct rupture geometries, and $e$ spans all rupture variations including events with the same rupture geometries but different slip distributions, thus the various rupture scenarios is a subset of the rupture variations ($\{l\} \subseteq \{e\}$).
The term $\bar{y}_{ls} = \frac{1}{|Y_{ls}|}\sum_{y_{es} \in Y_{ls}} y_{es}$ is the mean residual across all variations for rupture scenario $l$ at site $s$ 
The terms $\delta \dot{B}_l $ and $\delta \dot{W}_{ls}$ represent the aleatory variability of $\bar{y}_{ls}$ that exists between different rupture scenarios, and is not explainable by the non-ergodic path and source effects. This component is referred to as secondary aleatory variability in this paper. The terms $\delta \ddot{B}_e$ and $ \delta \ddot{W}_{es}$ describe the variability of $y_{es}$ about $\bar{y}_{ls}$. This variability results from the stochastic rupture generator used in CyberShake to characterize the randomness in rupture slip distribution when the rupture dimension, location, and geometry are fixed. This component is referred to as primary aleatory variability in this paper. 

The secondary aleatory variability ($\delta \dot{B}_l $ and $\delta \dot{W}_{ls}$) reflects the limited representational capacity of the GP model, and can, in principle, be reduced to zero if a "perfect" GP model is achieved. In contrast, the primary aleatory variability ($\delta \ddot{B}_e$ and $ \delta \ddot{W}_{es}$ ) is intrinsic to the stochastic rupture process and cannot be reduced within the NGMM framework.

Following the decomposition in Eq. \ref{eq:two_steps}, a GP model is developed for $\bar{y}_{ls}$ (see Eq. \ref{eq:GP_formulation_mean_1} - \ref{eq:GP_formulation_mean_3}) and a linear mixed-effects model (LMM)\cite{verbeke2000linear} is used to describe $\delta \ddot{B}_e$ and $\delta \ddot{W}_{es}$. LMMs are widely used to describe the variability of repeated measures while distinguishing individual-specific variability (e.g., within-event uncertainty in ground motion modeling) and global population-level variability (e.g., between-event uncertainty). In LMM, $\delta \ddot{B}_e$ and $\delta \ddot{W}_{es}$ are modeled as two multivariate normal with covariances $\ddot{\tau}^{2}\mathbf{1}$ and $\ddot{\phi}^{2}\mathbf{I}$, repectively. The parameters $\ddot{\tau}^{2}$ and $\ddot{\phi}^{2}$ can be estimated with a small computational cost (see Appendix \ref{sec:appendix_1}). This decomposition significantly improves computational efficiency. Specifically, the GP model is trained on $\bar{y}_{ls}$, which is approximately 84 times smaller (corresponding to the average number of rupture variations per scenario) than the full dataset $y_{es}$, thereby substantially reducing the computational burden.

When predicting earthquake scenarios not included in $\mathbf{y}_{\mathrm{obs}}$, $\delta \ddot{B}_e + \delta \ddot{W}_{es}$ follows the distribution $ \mathcal{N}(\mathbf{0}, \ddot{\tau}^{2}\mathbf{1} + \ddot{\phi}^{2}\mathbf{I})$ and is independent of  $\bar{y}_\mathrm{pred}$. Therefore, $\mathbf{y}_\mathrm{pred}$ is the sum of two independent multivariate normal variables, whose mean and covariance follow Eq. \ref{eq:GP_formualation_mean_pred}. The total secondary aleatory variability is $\delta \dot{B}_l  + \delta \dot{W}_{ls}$, and the total primary aleatory variability is $\delta \ddot{B}_e + \delta \ddot{W}_{es}$. 

When using the NGMM to interpolate partially observed earthquake scenarios, the terms $\delta \dot{B}_l$ and $\delta \ddot{B}_e$ can be inferred from observed ground motions in the same event, resulting in reduced aleatory variability than the prediction case. In this setting, the mean and covariance of $\mathbf{y}_{\mathrm{pred}}$ are given by Eq. \ref{eq:GP_formualation_mean_interpolate}, where $\mathbf{z}_f\in \mathbb{R}^{n_{\mathrm{pred}}\times n_{\mathrm{rupv}}}$ is the random effect design matrix for the prediction data points. $n_{\mathrm{pred}}$ is the number of prediction points, and $n_{\mathrm{rupv}}$ is the number of earthquake rupture variations in the observation dataset. Each row of $\mathbf{z}_f$ corresponds to one prediction data point, with a value of 1 in the column associated with the corresponding rupture variation and zeros elsewhere. Similarly, $\mathbf{Z}_f$ is the concatenation of a sequence of $\mathbf{z}_f$, which denotes the random effect design matrix for the observed data points. Because the matrix $[\ddot{\tau}^2\mathbf{1}+\ddot{\phi}^2\mathbf{I}]$ is blocked diagonal, its inverse can be computed efficiently by inverting each block individually, making the computational expanse of Eq. \ref{eq:GP_formualation_mean_interpolate} much smaller compared with the GP covariance inversion in in Eq. \ref{eq:GP_formulation}.

\begin{subequations} \label{eq:GP_formulation_mean}
\begin{align}
    & \begin{bmatrix}
        \bar{\mathbf{y}}_{\mathrm{obs}} \\
        \bar{\mathbf{y}}_{\mathrm{pred}}
        \end{bmatrix}
        \sim
        \mathcal{N}\!\left(
        \mathbf{0},
        \begin{bmatrix}
        \mathbf{K}_{f} + \dot{\phi^{2}}\mathbf{I} + \dot{\tau}^{2}\mathbf{1} & \mathbf{k}_{f} \\
        \mathbf{k}_{f}^{\mathsf{T}} & \mathbf{K}^{*}_{f}
        \end{bmatrix}
        \right) \label{eq:GP_formulation_mean_1}\\
    & \mathbb{E}\left[\bar{\mathbf{y}}_{\mathrm{pred}}|\bar{\mathbf{y}}_{\mathrm{obs}}\right] = \mathbf{k}_{f} \left[ \mathbf{K}_{f} + \dot{\phi}^{2}\mathbf{I} + \dot{\tau}^{2}\mathbf{1} \right]^{-1} \bar{\mathbf{y}}_{\mathrm{obs}} \label{eq:GP_formulation_mean_2}\\
    & \mathrm{COV}\left[\bar{\mathbf{y}}_{\mathrm{pred}}|\bar{\mathbf{y}}_{\mathrm{obs}}\right] = \mathbf{K}^{*}_{f} - \mathbf{k}_{f}^{\mathsf{T}} \left[ \mathbf{K}_{f} + \dot{\phi}^{2}\mathbf{I} + \dot{\tau}^{2}\mathbf{1} \right]^{-1}\mathbf{k}_{f} + \dot{\phi}^{2}\mathbf{I} + \dot{\tau}^{2}\mathbf{1} \label{eq:GP_formulation_mean_3} \\
    \text{For prediction\ \ \ \ }&\nonumber\\
    \begin{split}\label{eq:GP_formualation_mean_pred}
         & \mathbb{E}\left[\mathbf{y}_{\mathrm{pred}}|\mathbf{y}_{\mathrm{obs}}\right] = \mathbb{E}\left[\bar{\mathbf{y}}_{\mathrm{pred}}|\bar{\mathbf{y}}_{\mathrm{obs}}\right] \\
         & 
         \mathrm{COV}\left[\mathbf{y}_{\mathrm{pred}}|\mathbf{y}_{\mathrm{obs}}\right] =\mathrm{COV}\left[\bar{\mathbf{y}}_{\mathrm{pred}}|\bar{\mathbf{y}}_{\mathrm{obs}}\right] + \ddot{\tau}^{2}\mathbf{1} + \ddot{\phi}^{2}\mathbf{I}
    \end{split}\\
    \text{For interpolation}& \nonumber\\
    \begin{split} \label{eq:GP_formualation_mean_interpolate}
         & \mathbb{E}\left[\mathbf{y}_{\mathrm{pred}}|\mathbf{y}_{\mathrm{obs}}\right] = \mathbb{E}\left[\bar{\mathbf{y}}_{\mathrm{pred}}|\bar{\mathbf{y}}_{\mathrm{obs}}\right] + \ddot{\tau}^2\mathbf{z}_f\mathbf{Z}_f^T[\ddot{\tau}^2\mathbf{1}+\ddot{\phi}^2\mathbf{I}]^{-1}[\delta \ddot{B}_e + \delta \ddot{W}_{es}]_{\mathrm{obs}}\\
         & 
         \mathrm{COV}\left[\mathbf{y}_{\mathrm{pred}}|\mathbf{y}_{\mathrm{obs}}\right] =\mathrm{COV}\left[\bar{\mathbf{y}}_{\mathrm{pred}}|\bar{\mathbf{y}}_{\mathrm{obs}}\right] + \ddot{\tau}^2(\mathbf{z}_f\mathbf{z}_f^T - \mathbf{z}_f\mathbf{Z}_f^T[\ddot{\tau}^2\mathbf{1}+\ddot{\phi}^2\mathbf{I}]^{-1}\mathbf{Z}_f\mathbf{z}_f^T) + \ddot{\phi}^2\mathbf{I}
    \end{split}
\end{align}
\end{subequations} 

\subsection{Train-test split and hyperparameter tuning} \label{sec:tuning}
\begin{figure}
    \centering
    \includegraphics[width=0.5\linewidth]{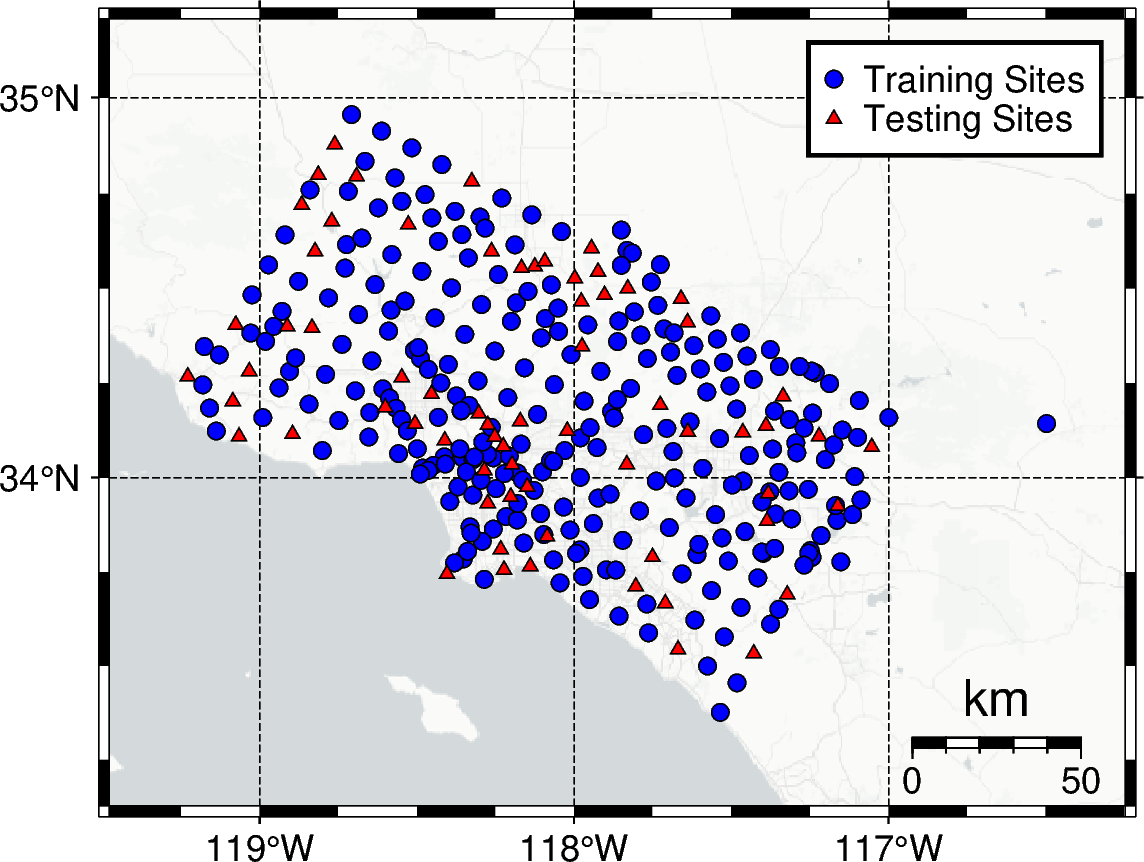}
    \caption{Sites in SECE CyberShake Study 22.12.}
    \label{fig:training_testing_sites_map}
\end{figure}

\begin{figure}
    \centering
    \includegraphics[width=0.85\linewidth]{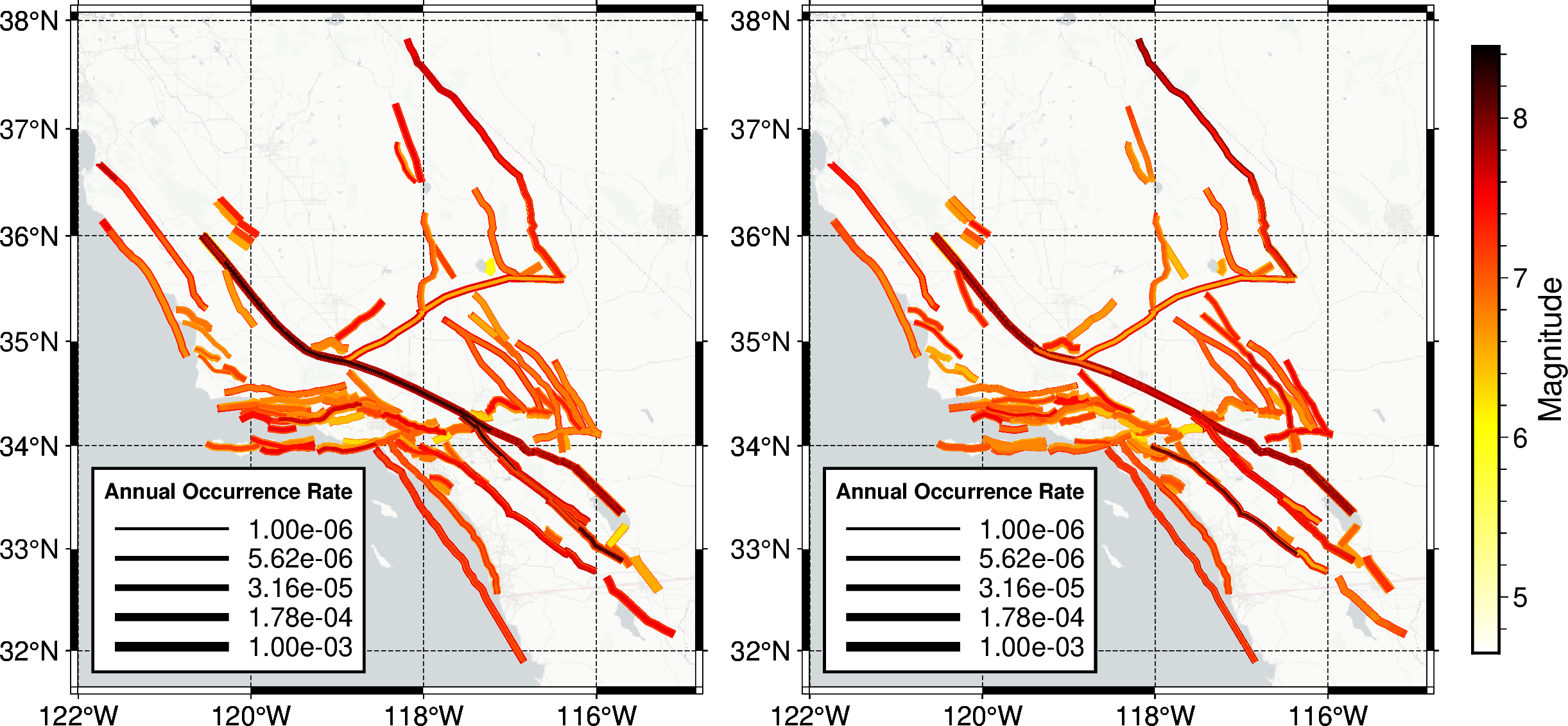}
    \caption{Earthquake scenarios used for NGMM development. Left: Training scenarios. Right: Testing scenarios.}
    \label{fig:training_testing_earthquakes_map}
\end{figure}

To evaluate the generalization performance of the NGMM to sites and earthquake scenarios not included in the training data, the CyberShake dataset is partitioned into training and test sets. This partitioning is enabled by the large volume of data available from the CyberShake simulations and distinguishes this study from most prior NGMM studies, where limited data necessitates using all observations for training, leaving no independent dataset for validation nor testing the models' generalization performance. As shown in Fig. \ref{fig:training_testing_sites_map}, the 335 CyberShake sites in Study 22.12 are randomly divided into 268 training sites and 67 testing sites, with approximately 25\% reserved for testing. Similarly, the earthquake scenarios 
are randomly split into two equal subsets of 4,179 training and 4,179 testing scenarios. A relatively large number of testing scenarios is intentionally retained to enable robust evaluation of the NGMM’s uncertainty-reduction capability in seismic hazard analysis, which requires a sufficiently large ensemble of earthquake scenarios. Consequently, a 50/50 split is adopted for earthquake scenarios. The spatial distribution of the earthquake scenarios, along with their magnitude and annual occurrence rate are shown in Fig. \ref{fig:training_testing_earthquakes_map}. 

By independently partitioning sites and earthquake scenarios, the dataset is divided into four groups: training earthquakes–training sites (TrTr), training earthquakes–testing sites (TrTe), testing earthquakes–training sites (TeTr), and testing earthquakes–testing sites (TeTe). Only the ground motion in the TrTr group is used to tune the GP kernel hyperparameters and serve as conditional observations in inference. The TrTe group is used to evaluate model performance in a data assimilation (interpolation) setting, where earthquake scenarios are partially observed. The TeTr and TeTe groups are used to assess predictive performance for earthquake scenarios not included in the training set. The number of data points in each group, along with the root-mean-squared error (RMSE) between CyberShake simulations and backbone ASK14 GMM predictions, are summarized in Table~\ref{table:train_test_split}.

\begin{table}
\caption{Train-test splitting and backbone GMM prediction error relative to CyberShake simulations.}
\label{table:train_test_split}
\centering
\small
\adjustbox{max width=\textwidth}{
\begin{tabular}{c c c c c}
\toprule
\multicolumn{1}{c}{Data group} &
\multicolumn{1}{c}{Number of} &
\multicolumn{1}{c}{RMSE } &
\multicolumn{1}{c}{RMSE } &
\multicolumn{1}{c}{Average StdDev*}
\\
\multicolumn{1}{c}{} &
\multicolumn{1}{c}{data points} &
\multicolumn{1}{c}{w.r.t. $\bar{y}$} &
\multicolumn{1}{c}{w.r.t. $y$}&
\multicolumn{1}{c}{} 
\\
\midrule
TrTr & 920,694 & 0.578 & 0.675 &  0.693\\
TrTe & 228,850 & 0.571 & 0.666 &  0.693\\
TeTr & 919,086 & 0.574 & 0.664 &  0.693\\
TeTe & 228,446 & 0.567 & 0.657 & 0.693 \\
\bottomrule
\end{tabular}
}
\normalsize
{\raggedright *The GMM provides a standard deviation (StdDev) for each prediction. This column reports the average predicted StdDev for each data group. \par}
\end{table}

The Python package GPyTorch \cite{gardner2018gpytorch} is employed to estimate the kernel hyperparameters using the maximum likelihood method. GPyTorch provides an efficient and modular implementation of GP models with GPU acceleration through its integration with PyTorch \cite{paszke2019pytorch}. Hyperparameter optimization requires repeated evaluation of the log-likelihood (Eq.~\ref{eq:loglikelihood}) and its gradients with respect to the hyperparameters. By leveraging automatic differentiation and GPU-accelerated linear algebra, GPyTorch enables efficient training on large datasets. Instead of the conventional marginal likelihood shown in Eq. \ref{eq:loglikelihood}, a leave-one-out cross-validation (LOO-CV) pseudo-likelihood objective (see section 5.4.2 of Williams and Rasmussen \cite{williams2006gaussian}) is used in this study. The optimization objective of LOO-CV is defined in Eq. \ref{eq:LOO-CV}, where the notation $\mathbf{y}_{-i}$ means all training data except data point $i$, with $\mathbb{E}\left[\mathbf{y}_{i}|\mathbf{y}_{-i}\right]$ and $\mathrm{COV}\left[\mathbf{y}_{i}|\mathbf{y}_{-i}\right]$ defined in Eq. \ref{eq:GP_formualation_mean_pred}, and $n$ is the number of data points. LOO-CV is generally more robust to outliers in the training data \cite{wahba1990spline}. 

\begin{subequations}\label{eq:LOO-CV}
    \begin{align} 
        \ln p\!\left(
\mathbf{y}_{i} \mid \mathbf{x}_{\mathrm{obs}}, \mathbf{y}_{-i}
\right)
&= -\frac{1}{2}\ln{\mathrm{COV}\left[\mathbf{y}_{i}|\mathbf{y}_{-i}\right]} - \frac{(y_{i} - \mathbb{E}\left[\mathbf{y}_{i}|\mathbf{y}_{-i}\right] )^2}{2\mathrm{COV}\left[\mathbf{y}_{i}|\mathbf{y}_{-i}\right]} - \mathrm{constant} \\
\mathcal{L}_{\mathrm{LOO-CV}} &= \sum_{i=1}^{n} \ln p\!\left(
\mathbf{y}_{i} \mid \mathbf{x}_{\mathrm{obs}}, \mathbf{y}_{-i}
\right)
    \end{align}
\end{subequations}
Evaluating the LOO-CV objective over the entire TrTr dataset would require forming the full covariance matrix $\mathbf{K}_f$, which would demand terabytes of memory and is therefore infeasible, even on multi-GPU systems. To address this, a mini-batching strategy is adopted. At each training epoch, the data are randomly shuffled, and a subset is used to approximate the LOO-CV objective at each optimization step. This approach has been demonstrated to be effective in Lavrentiadis et al. \cite{Lavrentiadis2024Groningen}. A batch size of 10,000 is used in this study, which corresponds to the maximum size that can be accommodated by the NVIDIA A100 GPU employed. The optimization is run for 250 epochs, after which both the LOO-CV objective and the hyperparameters are observed to have converged. In addition to tuning the hyperparameters using the full TrTr group, a reduced model is trained using only the first 400 earthquake scenarios from the TrTr group. This smaller dataset is intended to assess the sensitivity of NGMM performance to training data size and to emulate typical conditions of NGMMs trained on recorded ground motions, where only a limited number of earthquake events are available. The model trained on the full TrTr dataset is referred to as NGMM-1, while the model trained on the reduced dataset is referred to as NGMM-2.

\subsection{Kullback-Leibler (KL) minimization-based sparse Cholesky factorization}
As shown in Eq. \ref{eq:GP_formulation_2} and Eq. \ref{eq:GP_formulation_3}, GP inference requires inverting the matrix $\left[ \mathbf{K}_{f} + \phi_0^{2}\mathbf{I} + \tau_0^{2}\mathbf{1} \right]$, which is computationally prohibitive for the large training datasets considered in this study. To address this challenge, the Kullback–Leibler (KL) divergence minimization–based sparse Cholesky factorization (KLSC) method \cite{schafer2021sparse} is employed to reduce both computational time and memory requirements. The KLSC method leverages the screening effect in spatial statistics, which suggests that distant observations have negligible influence on the prediction, and converts dense Cholesky factorization to be approximated by a series of sparse, much smaller, and embarrassingly parallel matrix inversions. A brief overview of the method is provided below, and readers are referred to Sch\"{a}fer et al. \cite{schafer2021sparse} for full derivations, theoretical guarantees, and complexity analysis.

For a dense covariance matrix $\Theta \in \mathbb{R}^{I\times I}$, such as $\left[ \mathbf{K}_{f} + \phi_0^{2}\mathbf{I} + \tau_0^{2}\mathbf{1} \right]$ in this paper, which is constructed by evaluating the kernel functions for a set of locations $\{ x_i\}_{i\in I}$, the KLSC method first reorders the columns of $\Theta$ into a reverse-maximin ordering. The reverse-maximin ordering is constructed by selecting an initial point arbitrarily, and then iteratively choosing the next point as the one that maximizes the minimum distance to all previously selected points. During this process, a sparsity pattern is defined for each point. Specifically, to define the sparsity pattern of the $k$-th point, previously selected points within a distance $\rho l_k$ are retained, where $l_k$ is the minimum distance from the $k$-th point to the previously selected points, and $\rho$ is a tuning parameter (typically $\rho = 2$).

Given the sparse pattern associated with each column of $\Theta$, the KLSC method constructs a sparse lower-triangular matrix $\hat{L}$. The $k_{\mathrm{th}}$ column of $\hat{L}$ is calculated with Eq. \ref{eq:KLSC_column}, where $\Theta_{s_k,s_k}$ denotes the submatrix of $\Theta$ that only involves the sparse pattern of the $k_{\mathrm{th}}$ data point and $\mathbf{e}_1$ is a vector with the first entry equal to one and all other entries equal to zero.
It can be shown \cite{schafer2021sparse} that for the given sparsity pattern, the KL divergence between the normal distributions $\mathcal{N}(0,\Theta)$ and $\mathcal{N}\left(0,(\hat{L}\hat{L}^T)^{-1}\right)$ is minimized among all distributions with the same sparsity structure. In this sense, $\hat{L}$ provides an optimal sparse approximation to the Cholesky factorization of $\Theta$. 
 \begin{equation} \label{eq:KLSC_column}
L_k = \frac{\Theta^{-1}_{s_k,s_k}\mathbf{e}_1}{\sqrt{\mathbf{e}_1^T\Theta^{-1}_{s_k,s_k}\mathbf{e}_1}}     
 \end{equation}

When predictions are required at many locations, such as multiple sites and earthquake scenarios in seismic hazard analysis, Sch\"{a}fer et al. \cite{schafer2021sparse} proposed an approach to aggregate data points into groups, so that the Cholesky factorizations of the matrices $\Theta_{s_k,s_k}$ in Eq. \ref{eq:KLSC_column} can be reused for solving multiple columns in $\hat{L}$ at once. This aggregated approach is adopted in this study. Although the original work demonstrated that solving $L_k$ for each aggregated group can be computed in parallel, parallel computation is not implemented in the computer code published by them. Therefore, the code is extended in this study to enable Message Passing Interface (MPI)–based parallelism, further improving computational efficiency.

\section{Model performance}
\subsection{Inferring ground motions}
\begin{figure}
    \centering
    \includegraphics[width=0.75\linewidth]{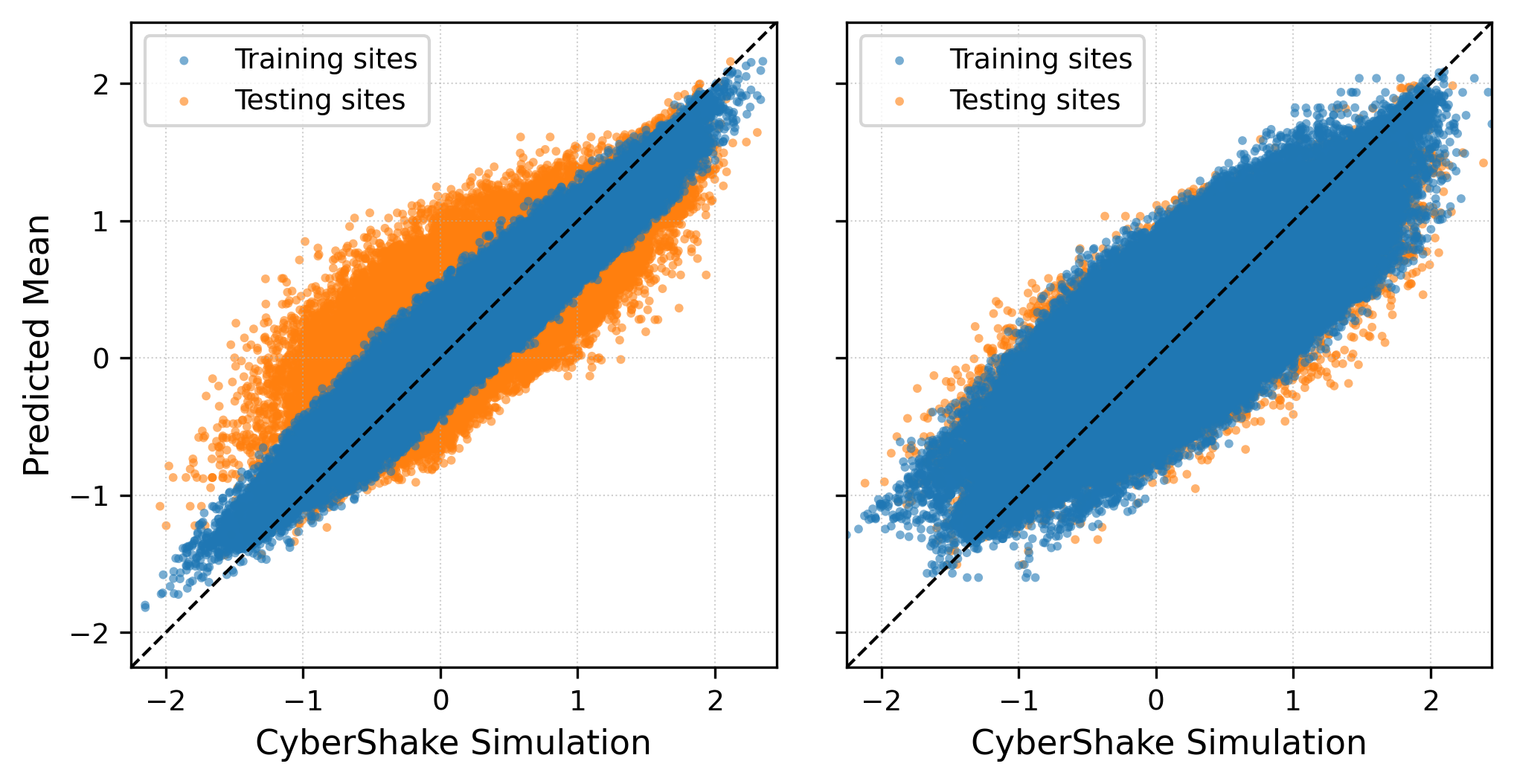}
    \caption{Comparison between NGMM prediction and CyberShake simulations. Left: Training earthquake scenarios. Right: Testing earthquake scenarios.}
    \label{fig:GP_accuracy}
\end{figure}

We first evaluate the performance of the trained NGMMs in inferring $\bar{y}$ and $y$. Following the procedures described in Sections~\ref{sec:collocation} and \ref{sec:tuning}, two NGMMs are developed using 400 (NGMM-2) and 4179 (NGMM-1) training earthquake scenarios from the TrTr group. The corresponding model
hyperparameters are listed in Table~\ref{table:hyperprameters}. Hyperparameter tuning requires approximately 2.5 and 11 node-hours on an NVIDIA A100 GPU for NGMM-2 and NGMM-1, respectively.

\begin{table}
\caption{Hyperparameter of the developed NGMM.}
\label{table:hyperprameters}
\centering
\adjustbox{max width=\textwidth}{
\begin{tabular}{c|c|cc|cc|cccc}
\toprule
                        Model&Number of training& \multicolumn{2}{c|}{$\delta P2P$} & \multicolumn{2}{c|}{$\delta S2S$} & \multirow{2}{*}{$\dot{\tau}^2$} & \multirow{2}{*}{$\dot{\phi}^2$} & \multirow{2}{*}{$\ddot{\tau}^2$} & \multirow{2}{*}{$\ddot{\phi}^2$} \\
                        name&scenarios& Length scale (km)  & Variance (StdDev) & Length scale (km)  & Variance (StdDev)  &                         &                         &                         &                       \\
                        \midrule
NGMM-1&4179 & 8.566             & 0.073 (0.270)   & 9.352              & 0.098 (0.313)    & 0.0360                                      & 0.0545                  & 0.0553                  & 0.0663   \\   
NGMM-2&400 & 6.173             & 0.070 (0.265)   & 11.351             & 0.177 (0.421)   & 0.0665                                      & 0.0461                  & 0.0567                  & 0.0665                \\
         \bottomrule
\end{tabular}
}
\end{table}

\begin{table}

\caption{Performance of the trained NGMM.}
\label{table:error}
\centering
\small
\adjustbox{max width=\textwidth}{
\begin{tabular}{c | c c c c c| c c c c c }
\toprule

\multicolumn{1}{c}{} &
\multicolumn{5}{|c}{NGMM-1 (4179 training scenarios)}&
\multicolumn{5}{|c}{NGMM-2 (400 training scenarios)}
\\
\midrule
\multicolumn{1}{c}{Data group} &
\multicolumn{1}{|c}{RMSE} &
\multicolumn{1}{c}{Reduction} &
\multicolumn{1}{c}{RMSE } &
\multicolumn{1}{c}{Reduction} &
\multicolumn{1}{c}{Average}&
\multicolumn{1}{|c}{RMSE } &
\multicolumn{1}{c}{Reduction} &
\multicolumn{1}{c}{RMSE } &
\multicolumn{1}{c}{Reduction} &
\multicolumn{1}{c}{Average}
\\
\multicolumn{1}{c}{} &
\multicolumn{1}{|c}{w.r.t. $\bar{y}$} &
\multicolumn{1}{c}{w.r.t. GMM} &
\multicolumn{1}{c}{w.r.t. $y$}&
\multicolumn{1}{c}{w.r.t. GMM*} &
\multicolumn{1}{c}{ StdDev**}&
\multicolumn{1}{|c}{w.r.t. $\bar{y}$} &
\multicolumn{1}{c}{w.r.t. GMM} &
\multicolumn{1}{c}{w.r.t. $y$}&
\multicolumn{1}{c}{w.r.t. GMM*} &
\multicolumn{1}{c}{StdDev**}
\\
\midrule
TrTr & 0.096 & 83.4\% & 0.279 & 58.7\% (61.9\%)& 0.281; 0.382 & 0.058 & 90.0\% &0.264 & 60.9\% (61.8\%)& 0.272; 0.374\\
TrTe & 0.262 & 54.1\% & 0.377 & 43.4\% (61.3\%)& 0.347; 0.439  & 0.205 & 64.1\% & 0.327 & 50.9\% (61.3\%)& 0.382; 0.459\\
TeTr & 0.197 & 65.7\% & 0.413 & 37.8\% (47.5\%) & 0.333; 0.482  & 0.244 & 57.5\%& 0.419 & 36.9\% (47.1\%)& 0.438; 0.563\\
TeTe & 0.275 & 51.5\% & 0.448 & 31.8\% (46.9\%)& 0.387; 0.522 & 0.309 &45.5\% & 0.459 & 30.1\% (46.6\%)& 0.489; 0.603\\
\bottomrule 
\end{tabular}
}
\normalsize
{\raggedright *Values in parentheses are the theoretically maximum achievable reduction, which does not depend on the number of training scenarios (i.e., the values are the same for 400 and 4179 training scenarios). \par}
{\raggedright **The first value is the standard deviation (StdDev) for $\bar{y}$ and the second value is the StdDev for $y$. \par}
\end{table}

The posterior mean $\mathbb{E}\left[\bar{\mathbf{y}}_{\mathrm{pred}}|\bar{\mathbf{y}}_{\mathrm{obs}}\right]$ is computed for both training and
testing earthquake scenarios using the two NGMMs. Fig. \ref{fig:GP_accuracy} shows a comparison between the NGMM-1 prediction and CyberShake simulations. The inference is computed using 10 HPC nodes, with a total computational cost of approximately 60 node-hours. The root mean squared errors (RMSE) of the four groups are summarized in Table. \ref{table:error}. For NGMM-1 (trained on 4179 scenarios), the model accurately recovers the $\bar{y}$ for the TrTr group, achieving an 83.4\% reduction in RMSE relative to the ergodic backbone model. When predicting ground motions at the training sites and testing earthquake scenarios (the TeTr group), the NGMM-1 achieved a 65.7\% reduction in RMSE compared to the backbone model. Although performance degrades when predicting at unseen sites (TrTe and TeTe groups), the NGMM still achieves more than a 50\% reduction in RMSE. Compared with the performance of predicting the $\bar{y}$ for the "TeTr" group, the RMSE for predicting the "TrTr" is much smaller. This difference is attributed to the reduction in secondary between-event variability ($\dot{\tau}^2$) when predicting the training earthquake scenarios, and the magnitude of the difference ($0.197-0.096=0.101$) is smaller but comparable to $\sqrt{\dot{\tau}^2} = 0.189$. Comparing the RMSEs of the TrTe and TeTe groups, the reduction in secondary between-event variability is less pronounced than the reduction for the training site groups. This suggests that inferring $\bar{y}$ for testing sites is harder than for training sites, where direct observations of site effects are available, and that it is easier for the GP to distinguish the non-ergodic effects from aleatory variability in the total residual.  
When trained on only 400 scenarios, NGMM-2 appears to outperform NGMM-1 on the TrTr and TrTe groups. This is because the training data sets are much smaller, and the model tends to memorize the data. When applied to the TeTr and TeTe groups, NGMM-2 performs worse than NGMM-1, but still significantly outperforms the backbone GMM. Overall, both NGMMs provide substantial improvements in predicting
$\bar{y}$ across observed and unobserved earthquakes, as well as observed and unobserved sites.

The uncertainty in $\bar{y}$ inference is quantified using the posterior standard deviation of the GP, computed as the square root of the diagonal entries of $\mathrm{COV}\left[\bar{\mathbf{y}}_{\mathrm{pred}}|\bar{\mathbf{y}}_{\mathrm{obs}}\right]$ (Eq. \ref{eq:GP_formulation_mean_3}). The average standard deviations for the four data groups are reported in Table~\ref{table:error}. In practice, the predicted standard deviation serves as a proxy for RMSE when ground truth is unavailable. It is observed that the GP tends to overestimate the standard deviation for inferring $\bar{y}$. This overestimation is likely because the aleatory variability variances ($\dot{\tau}^2$ and $\dot{\phi}^2$) are estimated from the full training dataset, whereas the inferences are evaluated on a subset of training data selected by the KLSC method. The KLSC method screens out weakly correlated training data, resulting in a smaller posterior variance than that estimated using Eq. \ref{eq:GP_formulation_mean_3}. Estimating posterior variance with noisy observations (i.e., aleatory variability) in sparse GP remains an open problem in spatial statistics \cite{schafer2021sparse,datta2016hierarchical}, and is beyond the scope of this paper. Although overestimation is significant for the TrTr group, it is moderate for the testing groups. Therefore, the overestimation is not critical in practice, as the NGMMs are not intended for inferring training data. Moreover, a slightly overestimated variance ensures a reasonable conservative hazard and risk estimates.

$\mathbb{E}\left[\mathbf{y}_{\mathrm{pred}}|\mathbf{y}_{\mathrm{obs}}\right]$ and $\mathbb{E}\left[\mathbf{y}_{\mathrm{intp}}|\mathbf{y}_{\mathrm{obs}}\right]$ are also evaluated for the testing and training earthquake groups (see Eq. \ref{eq:GP_formualation_mean_pred} and \ref{eq:GP_formualation_mean_interpolate}). For the TeTr and TeTe groups, the predicted $\mathbb{E}\left[\mathbf{y}_{\mathrm{pred}}|\mathbf{y}_{\mathrm{obs}}\right]$ is equal to $\mathbb{E}\left[\bar{\mathbf{y}}_{\mathrm{pred}}|\bar{\mathbf{y}}_{\mathrm{obs}}\right]$ and the RMSE with respect to $y$ is larger than that with respect to $\bar{y}$, due to the inclusion of primary aleatory variability of $y$ compared to $\bar{y}$. As discussed in section \ref{sec:collocation}, this aleatory variability can not be reduced within the NGMM framework, resulting in a minimum achievable prediction RMSE. This minimum achievable RMSE is approximately $\sqrt{\ddot{\tau}^2 + \ddot{\phi}^2} = 0.34$. The NGMM prediction achieved RMSE values of 0.413 and 0.448 for the TeTr and TeTe groups, respectively, which correspond to reductions of approximately 37.8\% and 31.8\% relative to the GMM. The values are also close to the maximum reduction an NGMM can theoretically achieve. For the TrTe groups, the NGMM achieved an RMSE of 0.377. This is close to the minimum achievable RMSE for interpolation, which is $\sqrt{\ddot{\phi}^2} = 0.257$. These results demonstrate the effectiveness of the NGMM in interpolating partially observed ground motion fields. Similar to the estimated $\mathrm{COV}\left[\bar{\mathbf{y}}_{\mathrm{pred}}|\bar{\mathbf{y}}_{\mathrm{obs}}\right]$, the estimated $\mathrm{COV}\left[\mathbf{y}_{\mathrm{pred}}|\mathbf{y}_{\mathrm{obs}}\right]$ and $\mathrm{COV}\left[\mathbf{y}_{\mathrm{intp}}|\mathbf{y}_{\mathrm{obs}}\right]$ (see the "Average StdDev" column in Table. \ref{table:error}) overestimate the RMSE, but they remained within an acceptable range. 

Comparing NGMM-1 and NGMM-2, increasing the number of training scenarios improves performance, but the gains are modest. This is because most of the remaining uncertainty lies in the primary aleatory variability ($\delta \ddot{B_e}$ and $\delta \ddot{W_{es}}$) which can not be reduced by increasing the size of the training dataset. This observation suggests that, using the NGMM framework, a small number of training earthquakes can considerably reduce ground-motion modeling uncertainty. Hence, the proposed framework has the potential to be applied to recorded ground-motion datasets. This observation also suggests that further reduction in ground-motion prediction uncertainty would require better understanding of and predictive ability for the variability associated with the earthquake rupture process.

\begin{figure}
    \centering
    \includegraphics[width=0.75\linewidth]{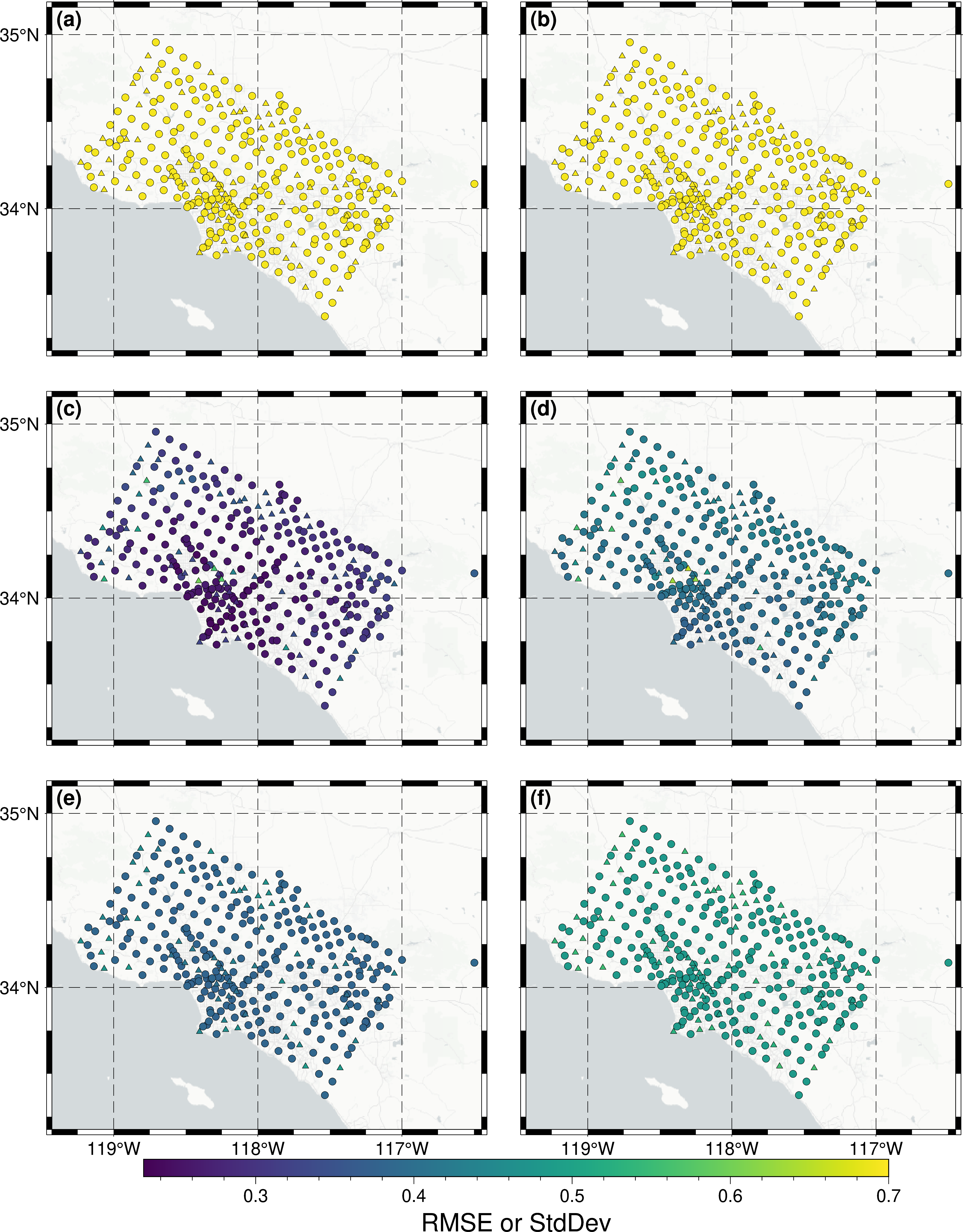}
    \caption{Spatial distribution of the root mean squared error (RMSE) and estimated standard deviation (StdDev). Circles represent training sites and triangles represent testing sites. (a,c,e) The RMSE and StdDev evaluated on the training earthquake scenarios; (b,d,f) The RMSE and StdDev evaluated on the testing earthquake scenarios; (a,b) StdDev estimated by the ergodic GMM; (c,d) RMSE of the NGMM-1 prediction; (e,f) StdDev estimated by the NGMM-1.}
    \label{fig:prediction_error_map_4179}
\end{figure}

Fig. \ref{fig:prediction_error_map_4179} presents the spatial distribution of each site's average RMSE and prediction standard deviation (StdDev). The three rows correspond to (i) the standard deviation from the ergodic GMM, (ii) the RMSE of NGMM redictions, and (iii) the posterior standard deviation from the NGMM. The left column shows results for training earthquake scenarios (TrTr and TrTe), while the right column shows results for testing scenarios (TeTr and TeTe). The plot shows that the NGMM achieves accurate ground-motion inference across the studied region, and the estimated model uncertainty (Fig. \ref{fig:prediction_error_map_4179}e and \ref{fig:prediction_error_map_4179}f) provides a reasonable upper bound on the actual prediction error (Fig. \ref{fig:prediction_error_map_4179}c and \ref{fig:prediction_error_map_4179}d). These results support the application of the NGMM to arbitrary locations within
the study area. A similar plot is produced for the NGMM-2 trained with 400 training scenarios (see Appendix \ref{sec:appendix_2}), which shows larger RMSE and StdDev than the NGMM-1. However, the uncertainty of the NGMM-2 remains substantially smaller than that of the GMM.

\subsection{Impact on seismic hazard}
\begin{figure}
    \centering
    \includegraphics[width=0.75\linewidth]{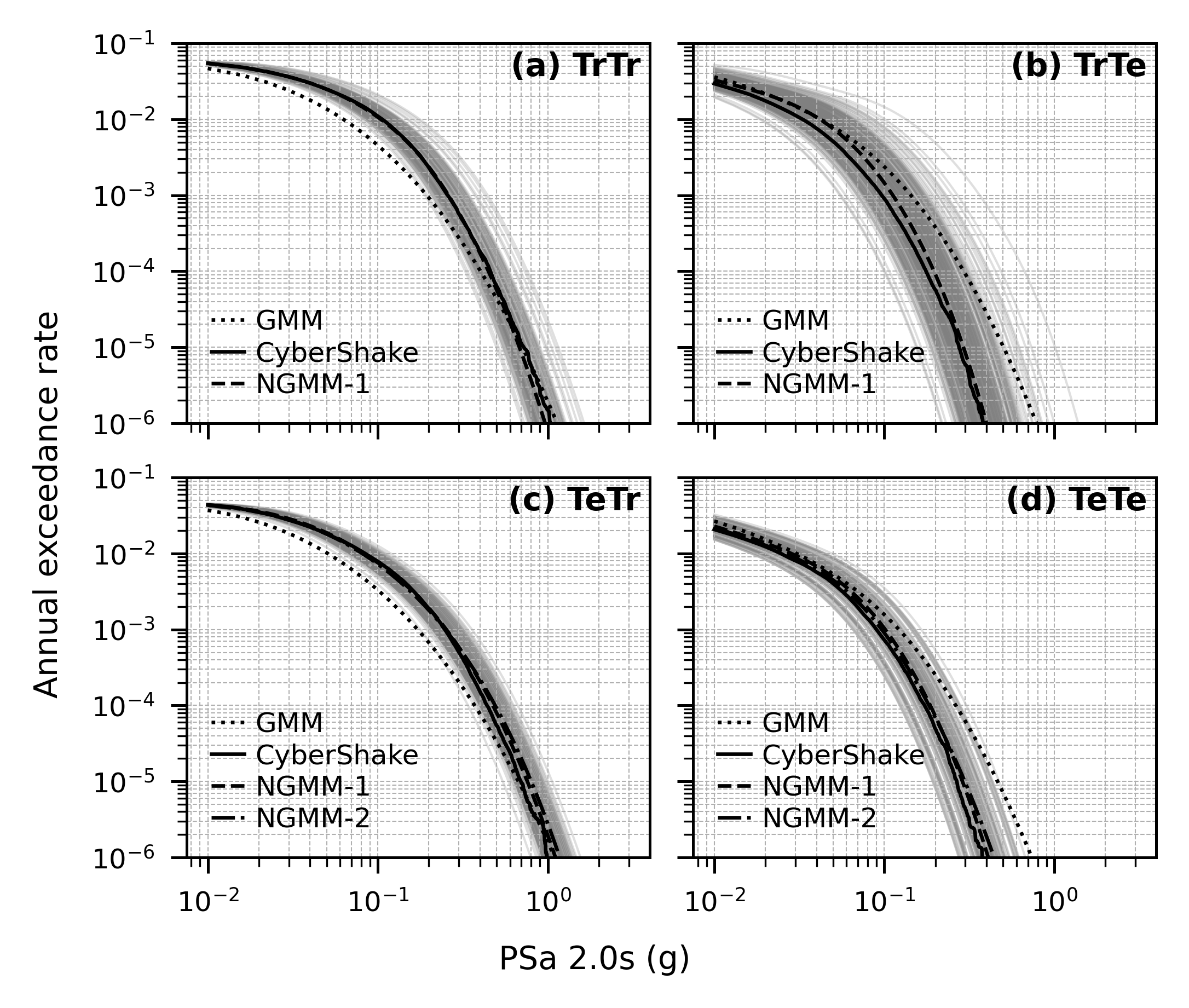}
    \caption{Hazard curves calculated with the ergodic GMM, CyberShake (ground truth), and the developed NGMMs. Grey curves represent samples from the NGMM-1's epistemic uncertainty (see Eq. \ref{eq:hazard_curve}). (a,c) Estimated hazards at the training site BRE; (b,d) Estimated hazards at the testing site MKBD; (a,b) Hazards evaluated using training earthquake scenarios; (c,d) Hazards evaluated using testing earthquake scenarios. }
    \label{fig:hazard_curves_MKBD_BRE}
\end{figure}

We further evaluate the performance of the NGMMs in seismic hazard estimation. Figure~\ref{fig:hazard_curves_MKBD_BRE} compares hazard curves computed using the NGMM, the backbone GMM, and CyberShake simulations for the four data groups. For the TrTr and TrTe groups, only NGMM-1 results are shown, as NGMM-2 is trained on only 400 earthquake scenarios and therefore does not provide comparable hazard estimates.

The hazard curves of GMM are calculated with Eq. \ref{eq:hazard_curve_a}, where $\mathbb{C}$ denotes the catalog of training or testing earthquake scenario, and $\lambda_i$ is the annual occurrence rate of the $i_{\mathrm{th}}$ scenario in the catalog. $\Phi$ is the cumulative density function (CDF) of a normal distribution with mean and standard deviation equal to $\bar{\mu}_i$ and $\bar{\sigma}_i$, both of which are deterministic in the GMM framework. For the NGMM, the hazard curves are calculated with Eq. \ref{eq:hazard_curve_b}, where $\mu_i$ and $\sigma_i$ are defined in Eq. \ref{eq:hazard_curve_NGMM}. In the NGMM formulation, $\mu_i$ is a random variable whose variance represents the epistemic uncertainty, while $\sigma_i$ represents the NGMM's aleatory variability. Therefore, Two approaches can be used to evaluate Eq.~\ref{eq:hazard_curve_b}. The first analytically integrates over the distribution of $\mu_i$, noting that a normal distribution with a normally distributed mean remains normal. The second approach draws samples of $\mu_i$ from Eq.~\ref{eq:hazard_curve_NGMM}, computes hazard curves for each realization, and then summarizes the results by taking the mean or median of the hazard curve realizations as the final hazard curve. The second approach is adopted in this paper, as it provides a clearer visualization of epistemic uncertainty and aleatory variability. 

The grey curves in Fig. \ref{fig:hazard_curves_MKBD_BRE} represent individual hazard curve realizations from NGMM-1, while the curves labeled as NGMM-1 and NGMM2 are the median of the corresponding samples. The hazard curves based on CyberShake simulations are computed with Eq. \ref{eq:hazard_curve_c}, where $\mathbb{R}_i$ is the catalog of rupture variations corresponding to the $i_{\mathrm{th}}$ rupture scenario, $|\mathbb{R}_i|$ represents the number of rupture variations in $\mathbb{R}_i$, and $\mathds{1}[\cdot]$ is the indicator function. The CyberShake curves are treated as the ground truth in this paper.


\begin{subequations} \label{eq:hazard_curve}
\begin{gather}
    \lambda_{GMM} (PSa > x) = \sum_{i\in \mathbb{C}} \lambda_i(1 - \Phi(\log(x); \bar{\mu}_i, \bar{\sigma}_i)) \label{eq:hazard_curve_a} \\
    \lambda_{NGMM} (PSa > x) = \sum_{i\in \mathbb{C}} \lambda_i(1 - \Phi(\log(x); \mu_i, \sigma_i)) \label{eq:hazard_curve_b} \\
    \lambda_{CyberShake}(PSa > x) = \sum_{i\in \mathbb{C}} \lambda_i\frac{1}{|\mathbb{R}_i|}\sum_{j\in \mathbb{R}_i} \mathds{1}[PSa_{ij} > x] \label{eq:hazard_curve_c}
\end{gather}
\end{subequations}

\begin{subequations} \label{eq:hazard_curve_NGMM}
\begin{align}
    \text{For prediction\ \ \ \ }&\nonumber\\
    \begin{split}\label{eq:hazard_curve_NGMM_a}
         & \mu_i \sim \mathcal{N}(\bar{\mu}_i+ \mathbb{E}\left[\mathbf{y}_{\mathrm{pred}}|\mathbf{y}_{\mathrm{obs}}\right], \mathbf{K}^{*}_{f} - \mathbf{k}_{f}^{\mathsf{T}} \left[ \mathbf{K}_{f} + \dot{\phi}^{2}\mathbf{I} + \dot{\tau}^{2}\mathbf{1} \right]^{-1}\mathbf{k}_{f}) \\
         & \sigma_i = \dot{\tau}^2+ \dot{\phi}^2 +\ddot{\tau}^2+ \ddot{\phi}^2
    \end{split}\\
    \text{For interpolation}& \nonumber\\
    \begin{split} \label{eq:hazard_curve_NGMM_b}
         & \mu_i \sim \mathcal{N}(\bar{\mu}_i+ \mathbb{E}\left[\mathbf{y}_{\mathrm{pred}}|\mathbf{y}_{\mathrm{obs}}\right], \mathbf{K}^{*}_{f} - \mathbf{k}_{f}^{\mathsf{T}} \left[ \mathbf{K}_{f} + \dot{\phi}^{2}\mathbf{I} + \dot{\tau}^{2}\mathbf{1} \right]^{-1}\mathbf{k}_{f} + \ddot{\tau}^2(\mathbf{z}_f\mathbf{z}_f^T - \mathbf{z}_f\mathbf{Z}_f^T[\ddot{\tau}^2\mathbf{1}+\ddot{\phi}^2\mathbf{I}]^{-1}\mathbf{Z}_f\mathbf{z}_f^T)) \\
         & \sigma_i = \dot{\phi}^2 + \ddot{\phi}^2
    \end{split}
\end{align}
\end{subequations} 

It can be observed that the NGMM curves provide a substantially closer approximation to the CyberShake curves, compared with the ergodic GMM curves. The agreement is strongest at
training sites (Fig. \ref{fig:hazard_curves_MKBD_BRE}a and \ref{fig:hazard_curves_MKBD_BRE}c), where site-specific effects are directly informed by observations. At testing sites (Figs.~\ref{fig:hazard_curves_MKBD_BRE}b and \ref{fig:hazard_curves_MKBD_BRE}d), the NGMM still performs well, although with slightly larger discrepancies due to the need to extrapolate site effects. Fig. \ref{fig:hazard_curves_MKBD_BRE}b illustrates the interpolation capability of the NGMM for estimating hazards at locations without direct simulations. The spread of NGMM curve realizations is wider than those in Fig. \ref{fig:hazard_curves_MKBD_BRE}d. This is because both the primary and secondary within-event variability are epistemic uncertainty in interpolation, while they are aleatory variability in prediction (see Eq. \ref{eq:hazard_curve_NGMM}). 
This within-event epistemic uncertainty can be reduced as more data becomes available. Quantitatively, using the maximum absolute error (Kolmogorov–Smirnov distance) between the model-predicted curves and the CyberShake curves for testing earthquakes suggested that the NGMMs achieved an average of 28\% error reduction across all sites, compared with the ergodic GMM. When evaluating the error using mean absolute error (integrating the absolute error over PSa), the NGMM achieved an average error reduction of 59\%. These results demonstrate the effectiveness of the NGMM in improving seismic hazard estimation relative to conventional ergodic GMMs.

\section{Application to power transmission infrastructure assessment}

We demonstrate the application of the developed NGMM through a seismic performance analysis of the power transmission network under a potential Mw~6.7 Puente Hills scenario. This scenario has been shown \cite{usgs_puentehills_scenario_2012,woods2017comparison} to pose a high loss potential because of its proximity to the densely populated Metropolitan Los Angeles area. Stochastic ground-motion fields are sampled using three approaches: the ergodic GMM, the NGMM as an interpolator, and the NGMM as a predictor. For each ground-motion realization, damage to transmission substations is estimated, and the resulting total unmet power demand is then
evaluated using a power-flow analysis tool, PyPSA \cite{brown2017pypsa}. By comparing the estimated population affected by power outages across the three ground-motion modeling approaches, we illustrate the value of the NGMM in reducing uncertainty in seismic performance analysis and reducing the  basis for decision-making.

\subsection{Exposure and direct physical damage modeling}
\begin{figure}
    \centering
    \includegraphics[width=0.55\linewidth]{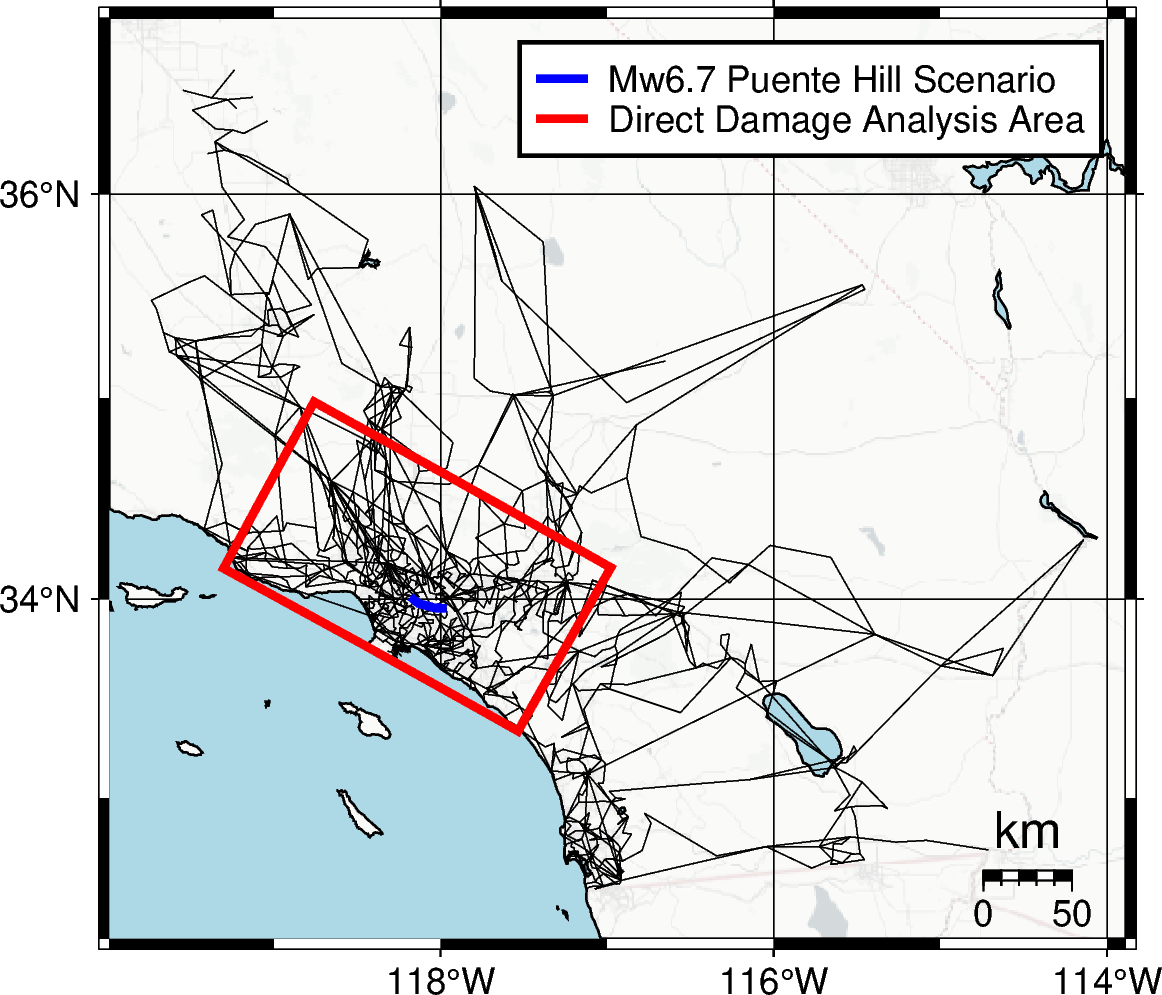}
    \caption{The studied power transmission network. Only the damage to the substations in the red box is modeled. }
    \label{fig:network_map}
\end{figure}

Exposure modeling in regional risk analysis refers to the process of identifying the physical assets and populations that may be affected by a hazard event. In this study, a synthetic power transmission network (Fig.~\ref{fig:network_map}) is used to represent the system in the study region. The network is extracted from an open-source national synthetic power transmission network model \cite{kunkolienkar2024description}, which was carefully developed as a realistic electric-grid model for research and educational purposes and has since been adopted in multiple power planning and operational studies
\cite{zhang2025short,jagadeesan2025optimal}. Although the network topology is synthetic and has relatively low granularity, the model is sufficient for this study, as the purpose is to demonstrate the benefit of a more accurate GMM in infrastructure performance evaluation, rather than to provide real policy-making or hazard-resilience modeling conclusions.

In this paper, we extract the portion of the network corresponding to the two Regional Energy Deployment System (ReEDS) zones in Southern California, resulting in a system with 1088 buses and 1559 lines. Building on the national synthetic power transmission network model \cite{kunkolienkar2024description}, the open-source project PyPSA-USA \cite{tehranchi2025pypsa} further integrated data sources, including detailed power generator and storage characteristics, power demand, fuel resources, and renewable policies, to enable flexible power-flow modeling and transmission expansion analysis. In PyPSA-USA, power supply from generators and storage devices is defined following data from the Public Utility Data Liberation \cite{selvans2021public} database, and the power demand at each bus is defined with either historical data or forecasted future demand. The forecasted demand from the National Renewable Energy Laboratory (NREL) Electrification Futures Study \cite{zhou2021electrification} for 2030 is used to represent the power demand in this paper. 

Direct physical damage in this study is modeled using the FEMA HAZUS methodology
\cite{fema_hazus_earthquake}. In PyPSA-USA, all buses in the studied region are represented as 230~kV substations, which are classified as medium substations in HAZUS. Given the high seismic hazard in the study area, all substations are assumed to be anchored. Under these assumptions, HAZUS provides fragility curves
that describe the probability of exceeding a series of damage states as a function of ground-motion intensity. The HAZUS fragility curves are lognormal cumulative density functions, as shown in Eq. \ref{eq:translate_IM_1}, where $DS_i$ is a random variable describing the damage state of the $i_{\mathrm{th}}$ substation and $ds_k$ stands for the $k_{\mathrm{th}}$ damage state (i.e., one of slight, moderate, extensive, and complete). $\Phi$ is the standard normal cumulative distribution function and $\alpha_k$ and $\beta_k$ are the median and logarithmic dispersions of the $k_{\mathrm{th}}$ curve. $\mathrm{PGA}_i$ is the peak ground acceleration (PGA) experienced by the $i_{\mathrm{th}}$ substation. 

The HAZUS fragility curves are defined in terms of PGA, whereas the NGMM developed in this paper applies to PSa 2.0~s. Therefore, the HAZUS fragility curves are translated to PSa 2.0~s using Eq. \ref{eq:translate_IM_2}. In Eq. \ref{eq:translate_IM_2}, $SD_{i, 2.0s}$ is the ASCE 7-22 design spectral PSa at 2.0~s for the $i_{\mathrm{th}}$ substation and $SD_{\mathrm{i,PGA}}$ is the corresponding design spectral PGA. Because the design spectral targets to achieve the same collapse risk across structures with different periods, this translation ensures that the long-term damage risk calculated using Eqs. \ref{eq:translate_IM_1} and \ref{eq:translate_IM_2} is roughly equivalent. Although the ASCE 7-22 design spectra were developed mainly for building structures, and equivalence in long-term risk does not imply exact equivalence of damage in a specific earthquake scenario, Eq. \ref{eq:translate_IM_2} is considered an adequate approximation in this study, as the purpose is to compare the effects of different ground-motion modeling approaches, and small inaccuracies in the fragility transformation do not alter that comparison.

\begin{subequations} \label{eq:translate_IM}
\begin{align}
P(DS_i > ds_k | \mathrm{PGA}_i) &=\Phi \left(\frac{\log(\mathrm{PGA}_i)-\log(\alpha_k)}{\beta_k} \right)\label{eq:translate_IM_1}\\
P(DS_i > ds_k | \mathrm{PSa}_i) &=\Phi \left(\frac{\log(\mathrm{PSa}_i)-\log(\alpha_k\frac{SD_{i, 2.0s}}{SD_{\mathrm{i,PGA}}})}{\beta_k} \right)\label{eq:translate_IM_2}        
\end{align}
\end{subequations}

\begin{figure}
    \centering
    \includegraphics[width=0.95\linewidth]{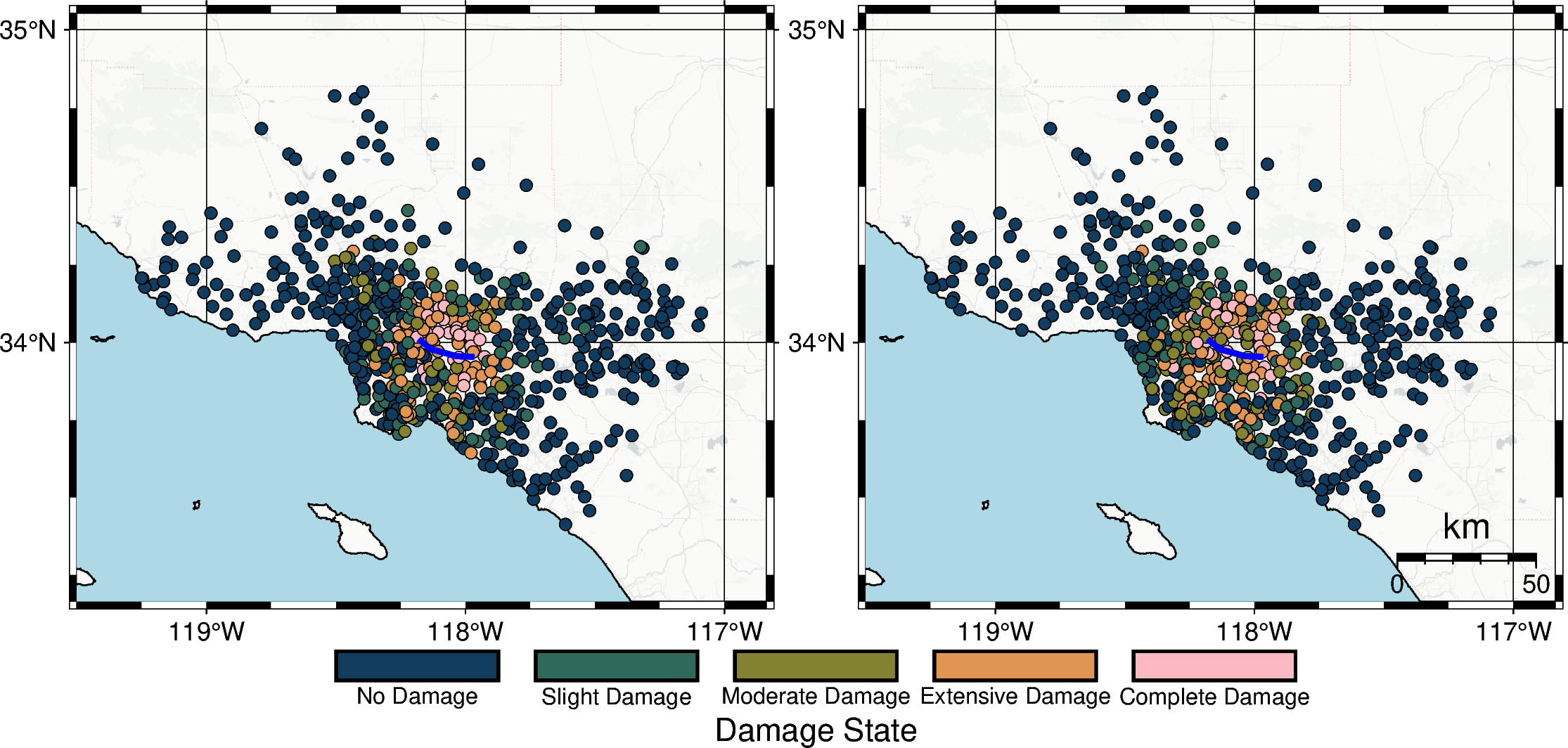}
    \caption{Two examples of substation damage state realizations. }
    \label{fig:substation_damage_state_spatial}
\end{figure}

\begin{figure}
    \centering
    \includegraphics[width=0.95\linewidth]{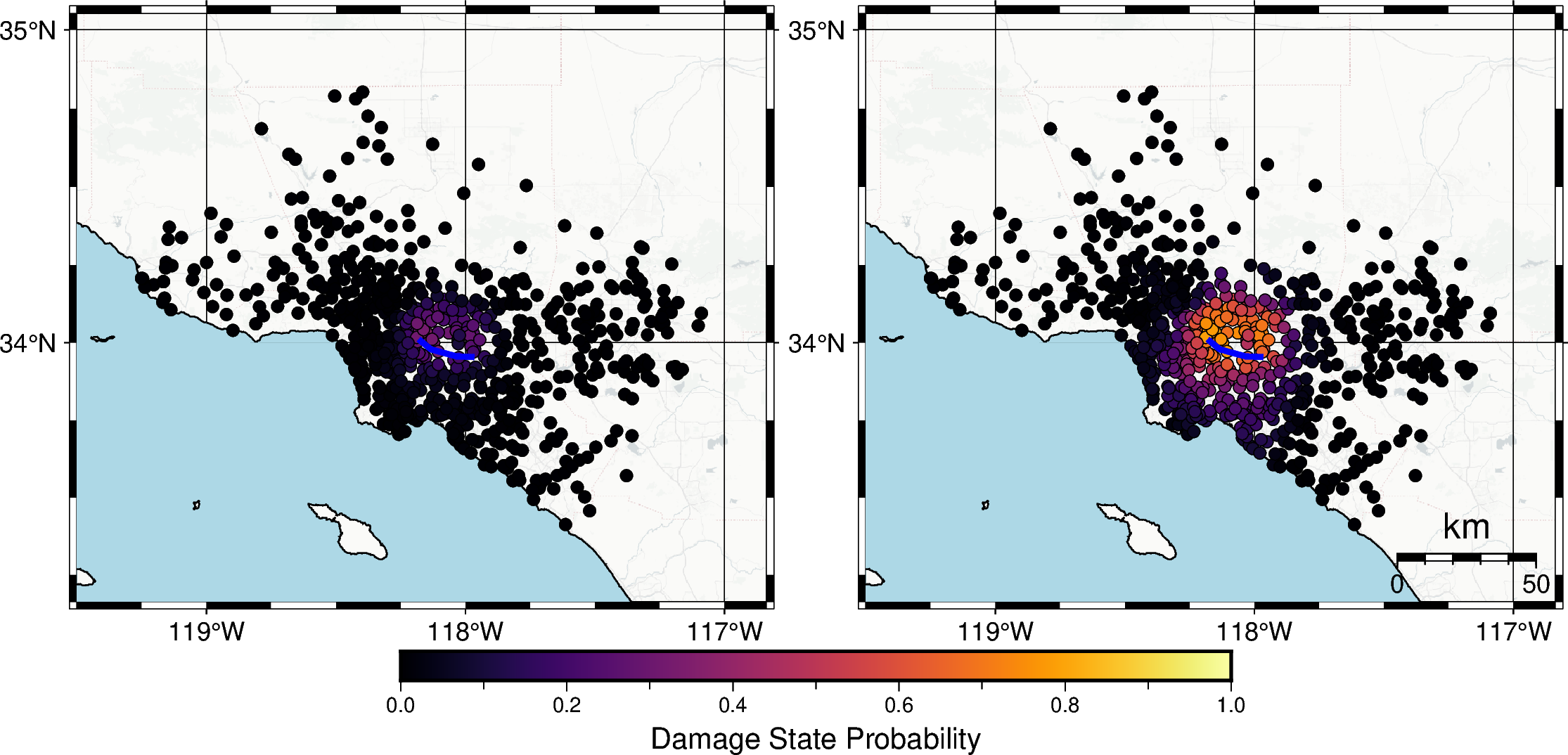}
    \caption{Probability of complete damage (left) and extensive (right) damage of each substation. }
    \label{fig:substation_damage_probability}
\end{figure}

To evaluate the performance of the power network, realizations of PSa 2.0~s ground-motion fields are first sampled using the median and covariance predictions of NGMM-1. For a given ground-motion realization, the damage state of each substation is modeled as a multinomial random variable with five possible outcomes, ranging from no damage to complete damage. The probability of each outcome is calculated from Eq.~\ref{eq:translate_IM_2}. Conditioned on the ground-motion field, substation damage states are assumed to be statistically independent in this study, and joint damage-state realizations are generated by sampling from the corresponding multinomial distributions. The Python package Pelicun \cite{zsarnoczay2020pelicun,zsarnoczay2025open} is extended in this paper to efficiently generate damage state realizations. A total of 1000 ground motion and damage state realizations are generated, and Fig. \ref{fig:substation_damage_state_spatial} shows two examples of the damage state realization maps. The damage probability at each substation is then estimated as the empirical frequency of each damage state across the 1000 realizations, and the corresponding maps for extensive and complete damage are shown in Fig.~\ref{fig:substation_damage_probability}. The expected number of completely damaged substations in this scenario is 23, estimated as the average over the 1000 realizations.

\subsection{System performance analysis}
\begin{figure}
    \centering
    \includegraphics[width=0.95\linewidth]{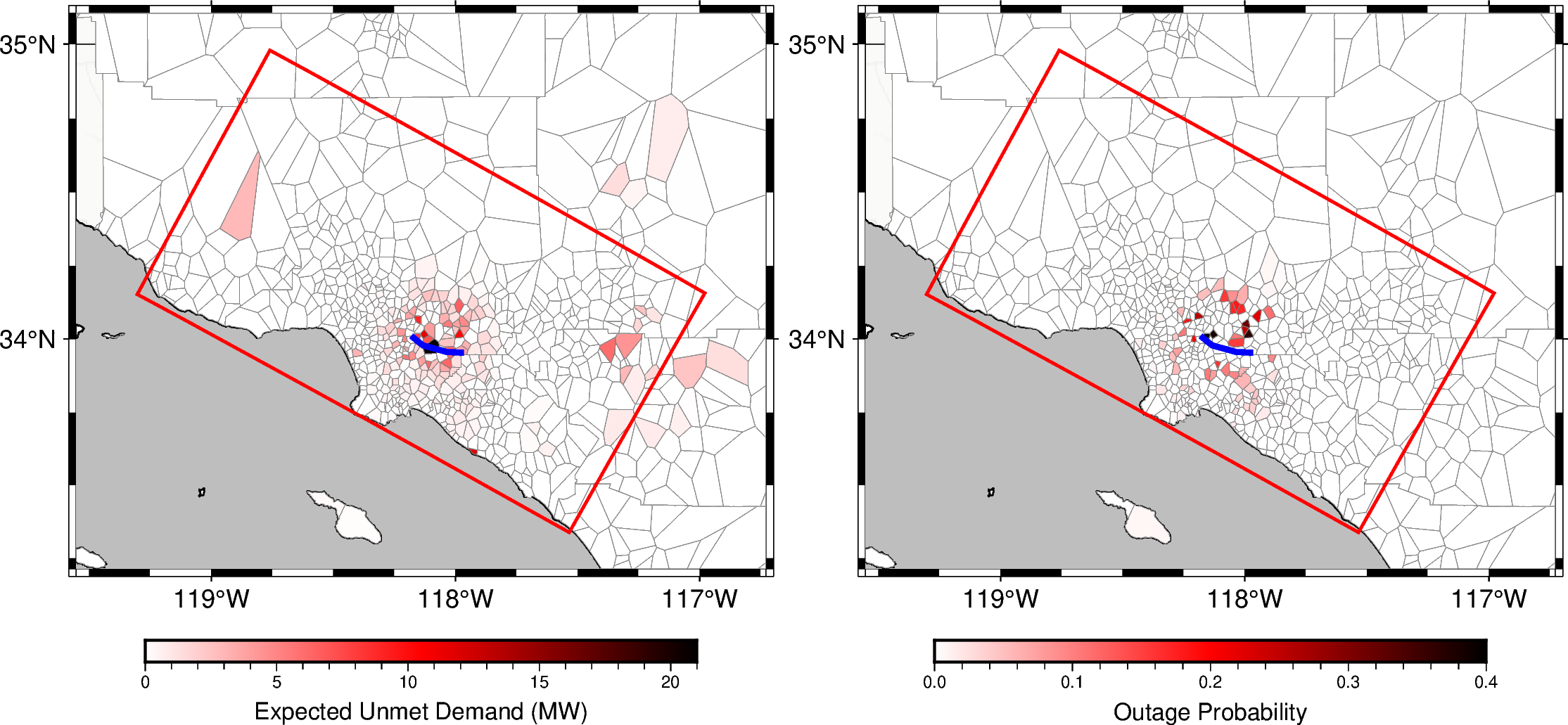}
    \caption{The spatial distribution of expected unmet power demand (left) and probability of power blackout (right) estimated with the NGMM. }
    \label{fig:pypsa_damage_spatial_ngmm_pred}
\end{figure}

Given a substation damage-state realization, such as those shown in Fig.~\ref{fig:substation_damage_state_spatial}, the residual functionality of each substation is estimated following the HAZUS methodology. The transmission ratings of lines connected to damaged substations are then reduced accordingly. Assuming power generators remain largely functional and that generation can be redispatched shortly after the earthquake, PyPSA-USA solves an optimization problem to determine the minimum-cost dispatch subject to power-line capacity constraints and Kirchhoff's voltage law. From this dispatch solution, the power supply at each bus is obtained, and the unmet demand in the service area of each bus is computed as the difference between demand and delivered supply. If the power supply at a bus drops to zero, the population in its service area is considered to experience a power outage. The service area of all buses and the population in each area are provided by the PyPSA-USA model. This analysis is repeated for all 1000 damage-state realizations, and Fig.~\ref{fig:pypsa_damage_spatial_ngmm_pred} shows the resulting expected unmet power demand (in MW) and the probability of power outage in the areas near the earthquake.

Although post-earthquake system operation is unlikely to exactly follow the optimal dispatch computed by PyPSA-USA, and the estimated unmet demand may therefore be approximate, the locations experiencing power outages are governed primarily by network connectivity and are expected to be similar to those identified by the model. For this reason, the population experiencing power outage is used here as the primary metric for comparing the system-performance results obtained with different ground-motion models. The system performance analysis is repeated four times using four ground motion modeling approaches: the NGMM-1 as a predictor, the ergodic backbone GMM as a predictor, the NGMM-1 as an interpolator, and a calibrated ergodic backbone GMM as a predictor. The calibrated ergodic backbone GMM is defined to have the same median as the NGMM-1, while the prediction standard deviation remains the same as the original backbone GMM, which is much larger than that of NGMM-1. As shown in Table. \ref{table:error}, the NGMM-1 achieves a high interpolation accuracy, and the corresponding performance analysis results are considered as a close approximation to those based on CyberShake simulations. The results obtained using the backbone GMM represent the conventional modeling approach, whereas those obtained using the calibrated ergodic GMM are intended to illustrate the effect of overestimating the true ground-motion variability on seismic risk estimation.

\begin{figure}
    \centering
    \includegraphics[width=0.5\linewidth]{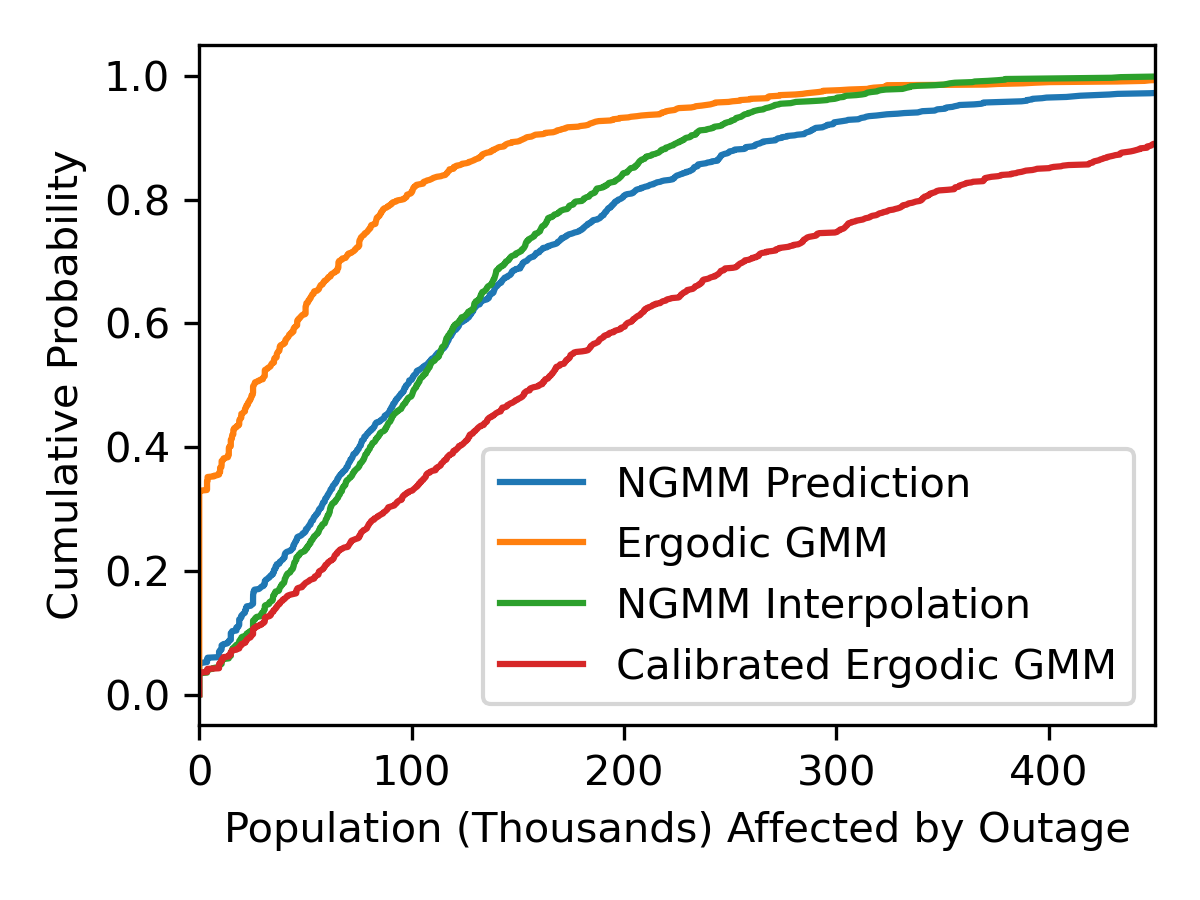}
    \caption{Cumulative probability density of the population affected by power outage estimated using four ground motion modeling approaches.}
    \label{fig:pypsa_damage_stats_comparison}
\end{figure}

For each ground-motion modeling approach, 1000 realizations of PSa 2.0~s fields are generated, and the corresponding outage populations are computed with PyPSA-USA. Fig. \ref{fig:pypsa_damage_stats_comparison} compares the cumulative probability density of the resulting power outage population. The NGMM prediction closely matches the interpolated CyberShake simulated ground motion, highlighting the accuracy of the developed NGMM for seismic risk modeling. In contrast, the ergodic backbone GMM substantially underestimates the outage population. This bias is caused by its underestimation of ground-shaking intensity in the study area (see Appendix \ref{sec:appendix_3} and Figure 9 of Graves et al. \cite{graves2011cybershake}). This underestimation is due to the limitations of the ergodic modeling framework in representing the three-dimensional basin effects \cite{graves2011cybershake,moschetti2024basin} near downtown Los Angeles, which amplify ground-shaking intensity. The developed NGMM model successfully captured the region-specific basin effect, leading to more accurate estimates of seismic performance. 

After calibrating the ergodic GMM to match the 
NGMM's median, the performance analysis substantially overestimates the outage population. This occurs because the calibrated ergodic GMM retains the much larger aleatory standard deviation of the original ergodic GMM, which leads to a higher probability of sampling extreme ground motions. Previous research \cite{Lavrentiadis2024Groningen} has shown that such large uncertainty can lead to overestimated seismic hazard estimates. Because system-performance metrics may increase nonlinearly with ground-shaking intensity, the impact of excessive uncertainty can be even more pronounced in seismic risk analysis, and this case study clearly demonstrates this effect. In other words, reducing uncertainty in ground motion modeling can reduce the seismic risk associated with aleatory variability  
, thereby reducing overly conservative seismic design. Avoiding overly conservative risk estimates can be particularly important for applications such as insurance pricing, and for reducing waste and improving the sustainability of the built environment.

\FloatBarrier

\section{Conclusions}\label{sec5}
This paper presents a non-ergodic ground motion model (NGMM) developed using the SCEC CyberShake Study 22.12 dataset, a large-scale physics-based simulation database comprising ground motions at 335 sites in the greater Los Angeles region across more than 770,000 rupture variations. The proposed NGMM employs a Gaussian Process (GP) framework to capture non-ergodic effects, together with a suite of scalability strategies that enable GP training and inference on multi-million–point datasets. Validation on unseen sites and earthquake scenarios, which are independent of training, demonstrates that the model can accurately interpolate partially observed ground-motion fields and predict ground motion for unobserved events. The applicability of the NGMM is demonstrated through a regional-scale performance assessment of a synthetic power transmission network in Southern California. Results show that neglecting non-ergodic effects, such as three-dimensional basin amplification, introduces substantial bias in infrastructure performance estimates. Moreover, reliance on ergodic ground motion models, due to their inflated aleatory variability, can lead to overestimation of seismic risk by as much as a factor of two for the evaluated infrastructure system.

A distinguishing feature of this work is that the data used by the NGMM consist of repeated rupture variations for the same rupture scenarios (e.g., geometry, magnitude, etc.), which enables direct characterization of aleatory variability associated with the inherent randomness in fault rupture processes. Accordingly, the proposed NGMM combines a GP model for the median ground motion conditioned on rupture characteristics and a linear mixed-effects model to represent within-event and between-event aleatory variability. To ensure computational tractability, a Kullback–Leibler (KL) divergence–based sparse Cholesky factorization is employed for covariance inversion, and the GPU-accelerated package GPytorch is adopted to enable efficient matrix algebra. These innovations collectively make NGMM training feasible for large physics-based simulation datasets that would otherwise be computationally prohibitive. The case study integrates ground motion modeling, component-level damage assessment based on FEMA HAZUS methodologies, and system-level functionality analysis using the open-source PyPSA-USA framework. To the authors’ knowledge, this is among the first end-to-end seismic performance assessments in the literature for power transmission networks at a regional scale.

Overall, the proposed NGMM framework enhanced the scalability of NGMMs and demonstrated the promising potential to improve ground motion model accuracy by incorporating emerging large ground motion data from advanced monitoring and simulations. The results highlight that improved ground-motion modeling can significantly reduce both bias and inflation in risk estimates due to excessive aleatory variability. The high accuracy of the NGMM trained on a smaller training data set suggests that the bottleneck in improving the NGMM lies in reducing the uncertainty associated with the fault rupture process, which is currently treated as aleatory variability and can not be reduced by increasing training data. Future research is needed to enhance the predictive capacity of fault rupture models and to integrate them with ground motion prediction frameworks.



\bmsection*{Data and Code Availability}

The computer code and data used to produce the analysis presented in this paper are available at \url{https://doi.org/10.17603/ds2-tksx-c844}.

\bmsection*{Acknowledgments}
The authors would like to acknowledge Scott Callaghan
and Philip J. Maechling for their support of the SCEC CyberShake dataset 

\bmsection*{Financial disclosure}
The authors are grateful for the financial support from the US Geological Survey, award No. G24AP00228, and the support from the NSF/Industry-University Collaborative Research Center "Geomechanics and Mitigation of Geohazards" (National Science Foundation, United States award No. 1822214)

\bmsection*{Conflict of interest}

The authors declare no potential conflict of interest.


\bmsection*{Supporting information}

Additional supporting information may be found in the
online version of the article at the publisher’s website.

\appendix

\bmsection{Linear mixed model for the primary aleatory variability}\label{sec:appendix_1}

The primary aleatory variability $\delta \ddot{B}_e + \delta \ddot{W}_{es}$ is modeled as a linear mixed model (LMM) in the NGMM formulation in this paper. In the LMM, the value of $\delta \ddot{B}_e$ is the same across all ground motions in the earthquake $e$, and it follows the normal distribution in Eq. \ref{eq:LMM_1}. The value of $\delta \ddot{W}_{es}$ is an independent realization of the normal distribution in Eq. \ref{eq:LMM_2} at any earthquake $e$ and site $s$.

\begin{subequations}
    \begin{align}
        \delta \ddot{B}_e &\sim \mathcal{N}(0, \ddot{\tau}^2) \label{eq:LMM_1}\\
        \delta \ddot{W}_{es} &\sim \mathcal{N}(0, \ddot{\phi}^2)\label{eq:LMM_2}
    \end{align}
\end{subequations}
where $\delta \ddot{W}_{es}$ is i.i.d. across each $s$ and $\delta \ddot{B}_e$ is independent of $\delta \ddot{W}_{es}$. The variance of $\ddot{\tau}^2$ and $\ddot{\phi}^2$ needs to be estimated from the training data. The maximum likelihood method is adopted in this paper to estimate these two values, and an efficient method for estimating the log-likelihood is described here. 

We first write the primary aleatory variability in the $e_{\mathrm{th}}$ earthquake in a vector form:
\begin{equation*}
    \delta \ddot{\mathbf{B}}_e + \delta \ddot{\mathbf{W}}_{es} = \delta \ddot{B}_e\mathbf{1}_{n_e} + \ddot{\mathbf{W}}_e
\end{equation*}
where $\mathbf{1}_{n_e}$ is a vector of $n_e$ ones, $n_e$ is the number of records in the $e_{\mathrm{th}}$ earthquake, $ \ddot{\mathbf{W}}_e\sim\mathcal{N}(0,\ddot{\phi}^2\mathbf{I}_{n_e})$, and $\mathbf{I}_{n_e}$ is an identity matrix with a dimension $n_e$. $\delta \ddot{\mathbf{B}}_e + \delta \ddot{\mathbf{W}}_{es}$ is the linear submission of two independent multivariate normal distributions. So the mean of $\delta \ddot{\mathbf{B}}_e + \delta \ddot{\mathbf{W}}_{es}$ is the sum of two multivariate normal distributions, which equals zero. The covariance of $\delta \ddot{\mathbf{B}}_e + \delta \ddot{\mathbf{W}}_{es}$ is $\mathbf{V}_e = \ddot{\tau}^2\mathbf{1}_{n_e}\mathbf{1}_{n_e}^\top + \ddot{\phi}^2\mathbf{I}_{n_e}$. 

Given the mean and covariance of $\delta \ddot{\mathbf{B}}_e + \delta \ddot{\mathbf{W}}_{es}$, the joint density for data in earthquake $e$ is:
\begin{equation*}
    p(\delta \ddot{\mathbf{B}}_e + \delta \ddot{\mathbf{W}}_{es} |\ddot{\tau}^2, \ddot{\phi}^2) = \frac{1}{(2\pi)^{n_e/2} \vert \mathbf{V}_e|^{1/2}} \exp\left( -\frac{1}{2}(\delta \ddot{\mathbf{B}}_e + \delta \ddot{\mathbf{W}}_{es})^\top \mathbf{V}_e^{-1}(\delta \ddot{\mathbf{B}}_e + \delta \ddot{\mathbf{W}}_{es})\right)
\end{equation*}
Taking the logarithm gives the log-likelihood of the data in the earthquake $e$ (up to a constant):
\begin{equation*}
    \ell_e (\ddot{\tau}^2, \ddot{\phi}^2) = \frac{1}{2} \left[\log|\mathbf{V}_e| +(\delta \ddot{\mathbf{B}}_e + \delta \ddot{\mathbf{W}}_{es})^\top \mathbf{V}_e^{-1}(\delta \ddot{\mathbf{B}}_e + \delta \ddot{\mathbf{W}}_{es})\right]
\end{equation*}
Estimating the log-likelihood requires estimating the determinant $|\mathbf{V}_e|$ and the quadratic form $(\delta \ddot{\mathbf{B}}_e + \delta \ddot{\mathbf{W}}_{es})^\top \mathbf{V}_e^{-1}(\delta \ddot{\mathbf{B}}_e + \delta \ddot{\mathbf{W}}_{es})$. To estimate the determinant, recall $\mathbf{V}_e = \ddot{\tau}^2\mathbf{1}_{n_e}\mathbf{1}_{n_e}^\top + \ddot{\phi}^2\mathbf{I}_{n_e}$ is a rank-1 update of $\ddot{\phi}^2\mathbf{I}$, so it has one eigenvalue $\ddot{\phi}^2+n_{e}\ddot{\tau}^2$ with an eigenvector $\mathbf{1}_{n_e}$ and $n_e-1$ eigenvalue $\ddot{\phi}^2$ with eigenvectors orthorgonal to $\mathbf{1}_{n_e}$. As the determinant of a matrix is the product of all eigenvalues, the value of $\log|\mathbf{V}_e|$ equals:
\begin{equation*}
    \log|\mathbf{V}_e|=\log(\ddot{\phi}^2 + n_e\ddot{\tau}^2) + (n_e-1)\log(\ddot{\phi}^2)
\end{equation*}

Now, define $\Lambda_e$ and $U$ as the eigen decomposition of $\mathbf{V}_e$:
$$U^\top\mathbf{V}_eU = \Lambda_e = \mathrm{diag}(\lambda_1, ...,\lambda_{n_e})$$ 

Also define $\mathbf{z}_e = U^\top(\delta \ddot{\mathbf{B}}_e + \delta \ddot{\mathbf{W}}_{es})$, the value of the quadratic form $(\delta \ddot{\mathbf{B}}_e + \delta \ddot{\mathbf{W}}_{es})^\top \mathbf{V}_e^{-1}(\delta \ddot{\mathbf{B}}_e + \delta \ddot{\mathbf{W}}_{es})$ is transformed as:
$$(\delta \ddot{\mathbf{B}}_e + \delta \ddot{\mathbf{W}}_{es})^\top \mathbf{V}_e^{-1}(\delta \ddot{\mathbf{B}}_e + \delta \ddot{\mathbf{W}}_{es})=\mathbf{z}^\top_e\Lambda_e^{-1}\mathbf{z}_e$$
which can be calculated row by row.

The first element of $\mathbf{z}^\top_e\Lambda_e^{-1}\mathbf{z}_e$ equals $\frac{z^2_{e,1}}{\ddot{\phi}^2+n_e\ddot{\tau}^2}$ and all other element equals $\frac{z^2_{e,k}}{\ddot{\phi}^2}$, where $z_{e,1}$ is the first element of $\mathbf{z}_e$ and $z_{e,k}$ is the $k_{\mathrm{th}}$ element of $\mathbf{z}_e$. So the quadratic term equals:
$$(\delta \ddot{\mathbf{B}}_e + \delta \ddot{\mathbf{W}}_{es})^\top \mathbf{V}_e^{-1}(\delta \ddot{\mathbf{B}}_e + \delta \ddot{\mathbf{W}}_{es})=\frac{z_{e,1}^2}{\ddot{\phi}^2+n_e\ddot{\tau}^2} + \sum_{k=2}^{n_e}\frac{z^2_{e,k}}{\ddot{\phi}^2}$$

Because the first column of $U^\top$ is $\frac{1}{\sqrt{n_e}}\mathbf{1}_{n_e}$, the value of $z_{e,1}$ is $\mu_e = \sum_{s=1}^{n_e}(\delta \ddot{B}_e + \delta \ddot{W}_{es}) $, which is the sum of data in earthquake $e$. Moreover, $\sum_{k=2}^{n_e}z^2_{e,k} = s_e^2 = \sum_{s=1}^{n_e}(\delta \ddot{B}_e + \delta \ddot{W}_{es}  - \frac{1}{n_e}\mu_e)$, which is the sample variance of the data in the earthquake $e$. So the quadratic form can be calculated from the sample mean and sample variance of the data in each earthquake, which are very cheap to compute.

Putting everything together, the log-likelihood for the data in earthquake $e$ is:
$$\ell_e (\ddot{\tau}^2, \ddot{\phi}^2) = -\frac{1}{2}\left[\log(\ddot{\phi}^2 + n_e\ddot{\tau}^2) + (n_e-1)\log(\ddot{\phi}^2) + \frac{\mu_e^2}{n_e(\ddot{\phi}^2+n_e\ddot{\tau}^2)+}+\frac{s_e^2}{\ddot{\phi}^2}\right]+\mathrm{const}$$

And the total log-likelihood is:
$$\ell (\ddot{\tau}^2, \ddot{\phi}^2) = \sum_e\ell_e (\ddot{\tau}^2, \ddot{\phi}^2)$$

The values of $\ddot{\tau}^2$ and  $\ddot{\phi}^2$ that maximize the total log-likelihood can then be estimated using optimization solvers with a low computation cost, and the "L-BFGS-B" optimizer available through the Python package SciPy\cite{virtanen2020scipy} is adopted in this paper.




\newpage
\FloatBarrier
\bmsection{Spatial distribution of error and uncertainty of NGMM-2}\label{sec:appendix_2}
\begin{figure}
    \centering
    \includegraphics[width=0.75\linewidth]{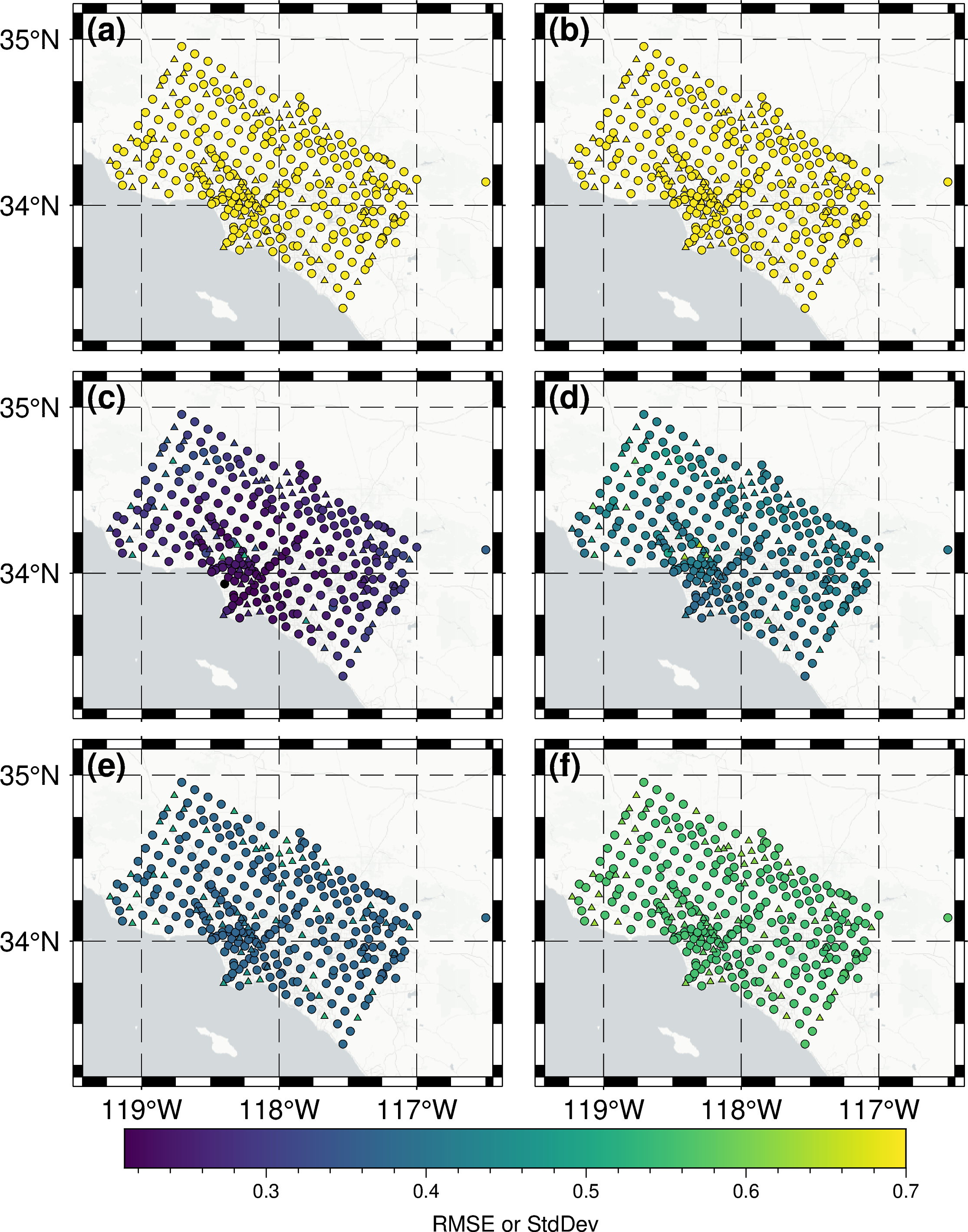}
    \caption{Spatial distribution of the root mean squared error (RMSE) and estimated standard deviation (StdDev). Circles represent training sites and triangles represent testing sites. (a,c,e) The RMSE and StdDev evaluated on the training earthquake scenarios; (b,d,f) The RMSE and StdDev evaluated on the testing earthquake scenarios; (a,b) StdDev estimated by the ergodic GMM; (c,d) RMSE of the NGMM-2 prediction; (e,f) StdDev estimated by the NGMM-2.}
    \label{fig:prediction_error_map_400}
\end{figure}


\FloatBarrier
\newpage
\bmsection{Ground motion prediction near downtown Los Angeles}\label{sec:appendix_3}%

\begin{figure}
    \centering
    \includegraphics[width=0.5\linewidth]{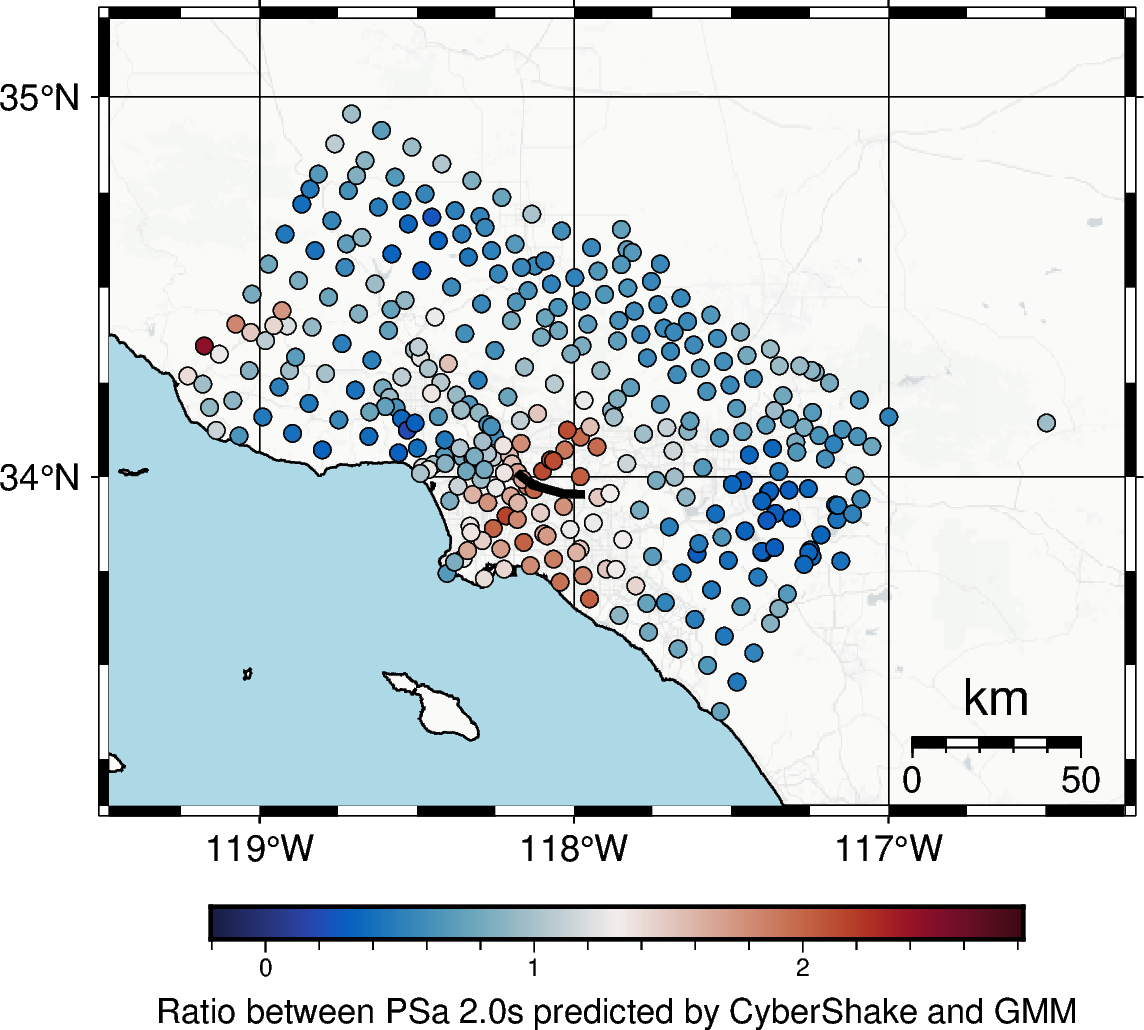}
    \caption{Ratio between the PSa 2.0s predicted by CyberShake and the ASK14 GMM. The red color near the studied Mw 6.7 earthquake scenario (black line in the figure) indicates ASK14's underestimation in ground motion.}
    \label{fig:cybershake_gmm_ratio}
\end{figure}







\end{document}